\documentclass[fleqn,usenatbib]{mnras}

\usepackage{newtxtext,newtxmath}
\usepackage[T1]{fontenc}

\DeclareRobustCommand{\VAN}[3]{#2}
\let\VANthebibliography\thebibliography
\def\thebibliography{\DeclareRobustCommand{\VAN}[3]{##3}\VANthebibliography}

\usepackage{graphicx}	
\usepackage{pdflscape}
\usepackage{booktabs}

\usepackage{lineno}
\usepackage{amsmath}	
\usepackage{xcolor}
\usepackage{transparent}
\usepackage{soul}
\usepackage{mathtools}
\usepackage{multirow}
\usepackage{hyperref}
\usepackage{subcaption}

\newcommand{\orcid}[1]{\href{https://orcid.org/#1}{\includegraphics[width=8pt]{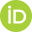}}}

\title[Stellar cycle effects on $\ion{He}{i}$~1083\,nm transit] 
{The effects of stellar activity cycles on planetary atmospheric escape and the $\ion{He}{i}$~1083\,nm transit signature} 
\author[A. P. Allan et al]{Andrew P. Allan\orcid{0000-0002-3900-5111},$^{1}$
Aline A. Vidotto\orcid{0000-0001-5371-2675},$^{1}$
Jorge Sanz-Forcada\orcid{0000-0002-1600-7835},$^{2}$
Carolina Villarreal D’Angelo\orcid{0000-0003-1701-7143}$^{3}$
 \\
 $^{1}$Leiden Observatory, Leiden University, P.O. Box 9513, 2300 RA Leiden,
 The Netherlands \\
$^{2}$Centro de Astrobiolog\'{i}a, CSIC-INTA, Camino bajo del Castillo s/n, 28692 Villanueva de la Ca\~nada, Madrid, Spain \\
$^{3}$Instituto de Astronom\'ia Te\'orica y Experimental. Laprida 854, X5000BGR Córdoba, Argentina 
 }

\date{Accepted 2025 October 21. Received 2025 October 20; in original form 2025 August 12}

\begin{document}
\label{firstpage}
\pagerange{\pageref{firstpage}--\pageref{lastpage}}
\maketitle

\begin{abstract}
The $\ion{He}{i}$~1083\,nm transit signature is commonly used in tracing escaping planetary atmospheres. However, it can be affected by stellar activity, complicating detections and interpretations of atmospheric escape. We model how stellar activity cycles affect the atmospheric escape and $\ion{He}{i}$~1083\,nm signatures of four types of highly irradiated exoplanets, at 0.025 and 0.05 au, during minimum and maximum cycle phases. We consider two stars, exhibiting different  cycle behaviours: the Sun and the more active star $\iota$~Hor, for which we reconstruct its spectral energy distributions at minimum and maximum phases using X-ray observations and photospheric models. We show that over a modulated activity cycle, the release of extreme ultraviolet photons, responsible for atmospheric escape, varies substantially more than that of mid-UV photons, capable of photoionising \ion{He}{I}~($2^3$S). This leads to consistently stronger helium signatures during maximum phases. 
We show that planets at the largest orbit are more affected by cycles, showing larger variations in escape rates and absorptions between minimum and maximum. 
We also confirm the counter-intuitive behaviour that, despite the fall-off in escape rate with orbital distance, the $\ion{He}{i}$~1083\,nm absorption is not significantly weaker at further orbits, even strengthening with orbital distance for some $\iota$~Hor planets.
We partially explain this behaviour with the lower mid-UV fluxes at more distant orbits, leading to less \ion{He}{I}~($2^3$S) photoionisations. 
Finally, we propose that stellar cycles could explain some of the conflicting $\ion{He}{i}$~1083\,nm observations of the same planet, with detections more likely during a phase of activity maximum.
\end{abstract}


\section{Introduction} 
\label{sec:intro}
The 1083\,nm helium triplet (hereby $\ion{He}{i}$~1083\,nm) signature, comprising of three individual lines centred on 1082.909, 1083.025, and 1083.034\,nm in air, has fast become the most popular method for tracing escaping atmospheres of highly irradiated exoplanets \citep[for recent analyses of the various detections and non-detections see][]{Orell-Miquel_2024_MOPYS, Krishnamurthy_Cowan_2024, Sanz-Forcada_2025_HeI_10830, McCreery_2025}. Its near-infrared (nIR) wavelength benefits from being observable from the ground, a strong advantage over hydrogen's Lyman-$\alpha$ line in the ultraviolet (UV), which is severely affected by absorption in the interstellar medium and requires space-based observations currently only possible with the Hubble Space Telescope \citep{Dos_Santos_2022_iauga_obs_of_pl_winds_outflows}. The $\ion{He}{i}$~1083\,nm triplet is produced by transitions from the 2$^3$S to the 2$^3$P state of neutral helium. 
The 2$^3$S state is metastable as radiative decays to the ground (1$^1$S) state are forbidden and hence greatly suppressed with a lifetime of 2.2 hours \citep{Drake1971}.

Despite the success of the $\ion{He}{i}$~1083\,nm feature in studying escaping planetary atmospheres, stellar activity can also affect the observed signature, sometimes making it difficult to distinguish planetary from stellar contributions. Planets for which stellar activity has been proposed to influence their $\ion{He}{i}$~1083\,nm observations include the sub-Neptune TOI-2076b \citep{Gaidos_2023_he_search_Two_200_Myr_old_Planets_TOI1807_TOI2076}, hot-Jupiter HAT-P-32b \citep{Zhoujian_Zhang_2023_HATP32b} and HD~189733b \citep{Salz_2018_hE_HD189733, 2020_Guilluy_GAPS_He_189, Zhang_Cauley_et_al_2022_variable_he_abs_hd189733b}. 
Various forms of stellar activity can affect $\ion{He}{i}$~1083\,nm either through direct $\ion{He}{i}$($2^3$S) variations in the stellar chromosphere \citep{Zirin_1975, Sanz-Forcada_Dupree_2008_active_cool_stars_he_1083} or indirect variations of $\ion{He}{i}$($2^3$S) material in the escaping planetary atmosphere transiting across the stellar disk. 
On direct variations due to the star, \citet{Mercier_2025_var_He_triplet_detection_lims_evap_atm} investigated temporal variability in the solar $\ion{He}{i}$~1083\,nm signature using multi-epoch observations, finding significant variability on timescales of minutes to days associated to telluric contamination and short-term stellar activity. They demonstrate that this observed solar $\ion{He}{i}$~1083\,nm variability should not significantly affect the inferred atmospheric escape properties of a strong 5 per cent planetary absorption transit signature. \citet{krolikowski2024strength} performed a similar temporal study of stellar $\ion{He}{i}$~1083\,nm variability for 10 young stars (with observational baselines ranging from minutes to years), cautioning that stronger variability in stars younger than 120\,Myr, would likely preclude even significant planetary transit signatures without well-timed out-of-transit reference observations. In this work, we will focus on indirect variations of $\ion{He}{i}$($2^3$S) material coming from the escaping planetary atmosphere transiting across the stellar disk.

Stellar activity operates on three main timescales. 
In the shorter of these timescales, stellar flares operate on the order of hours to days. For observations of the star alone, these bursts of electromagnetic radiation can cause variations in the stellar $\ion{He}{i}$~1083\,nm, as has been found in $\ion{He}{i}$~1083\,nm observations of the Sun \citep{Kuckein_2015_solar_flare} and M dwarf stars \citep{Fuhrmeister_2020_Carmenes_Mdwarfs_variability_triplet}. $\ion{He}{i}$~1083\,nm observations performed during a planetary transit are also susceptible to variations from stellar flares. For example,
\citet{2021_Vissapragada_search_He_in_V1298_tau_system} observed a flare over six high resolution spectra of V1298~Tau obtained with Habitable-zone Planet Finder \citep[HPF][]{Mahadevan_2012_HPF}. This flare coincided with a planetary transit of V1298~Tau~c, with the flare decay phase corresponding to an increase in the EW of the $\ion{He}{i}$~1083\,nm signature. However, they concluded that the enhancement was likely due to an increased population of \ion{He}{I}~($2^3$S) in the stellar chromosphere rather than of a planetary origin. Supporting this interpretation, they were unable to detect the transit of V1298~Tau~c at all using narrowband photometry centred on $\ion{He}{i}$~1083\,nm. A later higher-resolution non-detection of a transiting $\ion{He}{i}$~1083\,nm signature for V1298~Tau~c \citep{Alam_2024_young_non_detects} further supports this. However, 3-D hydrodynamic modelling performed for the hot-Jupiter WASP-69b, demonstrated that a stellar flare is also capable of enhancing the observed $\ion{He}{i}$~1083\,nm transit signature indirectly, through raising the planet's atmospheric escape \citep{Wang_Dai2021_WASP-69b}. Their modelled $\ion{He}{i}$~1083\,nm transit signature was temporarily weakened by an injected flare due to increased
\ion{He}{I}~($2^3$S) photoionisation before being enhanced after a few hours by which time the system had adjusted to a higher rate of atmospheric escape. 
Stellar activity in the form of stellar spots and bright facular or plage regions, also introducing short timescale variabilities, were shown to only weakly affect the $\ion{He}{i}$~1083\,nm transit signature, 
and is more likely to dilute rather than enhance it 
\citep{Cauley_2018_effects_stellar_activity_exo_transit}. Hence, stellar spots, bright facular and plage regions should be less problematic to escaping helium planetary studies than other forms of stellar activity which can greatly influence the $\ion{He}{i}$~1083\,nm transit signature.

Over the full lifetime of a highly irradiated exoplanet, a much longer timescale for stellar activity, planetary atmospheric escape is known to vary substantially. As the system ages and the star spins down \citep{Vidotto_2014}, the planetary atmospheric escape declines \citep{Owen_EVOL_ATM_ESCAP_REVIEW_2019, Allan_Vidotto_2019}. This is due to the reducing XUV flux responsible for heating the atmosphere \citep{Johnstone_2021} and the shrinking planetary radius \citep{Fortney_and_Nettelmann2010}. Consequently, the $\ion{He}{i}$~1083\,nm transit signature weakens over the evolution of the highly irradiated exoplanet, as shown previously in \citet{Allan_et_al_He_evol_2024}.

An intermediary timescale between the mentioned short-term (hours to days) and long-term (planetary lifetime) timescales is also worth investigating, namely that related to stellar activity cycles. The Sun exhibits an activity cycle which causes its emitted XUV flux to vary cyclically, with a period of 11 years \citep[][]{2010LRSP....7....1H}. As will be discussed further in section \ref{sec:XUV_cycle_general}, XUV cycles of other stars can differ substantially from this.
\citet{Hazra_2020} previously showed, using solar XUV observations, that solar-like cyclic XUV variations can influence the atmospheric escape of hot-Jupiters. They demonstrated that variations in the strength of atmospheric escape introduces a cyclic nature to the hydrogen Lyman-$\alpha$ and H-$\alpha$ tracers of atmospheric escape, with H-$\alpha$ being more sensitive to the stellar activity cycle. 
Similarly, \citet{Taylor_Koskinen_He_model_applied_to_hd209_2025} recently demonstrated a sensitivity of the $\ion{He}{i}$~1083\,nm signature of the hot-Jupiter HD209458b to a Sun-like activity cycle also using solar XUV data. The effect of a Sun-like stellar activity cycle on the $\ion{He}{i}$~1083\,nm transit signature has however yet to be explored for an exoplanet beyond the archetypal hot-Jupiter HD209458b. Furthermore, how the signature is affected by XUV cycles beyond that of our Sun is another interesting avenue worth exploring (see section \ref{sec:results_iota_hor}).

In this current work, we consider how stellar activity cycles affects the helium triplet signature of four different types of highly irradiated exoplanets at two orbital distances. In order to incorporate a stellar activity cycle, we take two different approaches for determining the stellar spectral energy distributions during minimum and maximum activity phases. The first approach described in section \ref{sec:XUV_cycle_Sun} considers a Sun-like activity cycle while our second approach considers the shorter cycle of $\iota$~Hor (section \ref{sec:SED_iHor}). Our method for modelling atmospheric escape and the $\ion{He}{i}$~1083\,nm signature using the 1-dimensional model of \citet{Allan_Vidotto_2025_young_He} is summarised in section \ref{sec:modelling_planets}. Our results are presented in section \ref{sec:results} along with relevant discussions in section \ref{sec:discussion}. 
\section{Stellar XUV activity cycles}
Although the activity cycle is better studied on the Sun, activity cycles also occur for other stars as made evident by monitoring programs such as that by the Mount Wilson \citep{Wilson_1978_mount_wilson_project, Baliunas_1995_mount_wilson_project2} and Lowell \citep{Hall_2007_Lowell_observatory} observatories. \citet{Sudeshna_Boro_Saikia_2018_chromospheric_activity_catalogue} more recently created a catalogue of 4454 cool stars exhibiting a cyclic nature using a combination of data from various surveys. The mentioned works all utilised the \ion{Ca}{ii} H\&K lines with visible wavelengths (in air) of 3968.469\,\AA\ and 3933.663\,\AA\ respectively, both indicators of chromospheric activity. However, cycles have also been detected at X-ray wavelengths for a handful of stars: 
 $\iota$~Hor \citep[][see section \ref{sec:SED_iHor} here]{Sanz_Forcada_2013_EARLIER_iota_hor_xray_cycle, Sanz_Forcada_2019_iota_hor_xray_cycle}, 
 {\ensuremath{\tau}}~Boo \citep{tau_boo_xray_cycle2017}, 
 HD 81809 \citep{Favata_2008_HD81809_xray_cycle, Orlando_2017_HD81809_xray_cycle}, 
 $\alpha$~Cen A \& B \citep{Robrade_2012_xray_61CygA_and_alpha_Cen_AB, Ayres2020_Alpha_Centauri_AandB_Xray_cycle}, 
 AB~Dor \citep{Singh_2024_AB_Dor_xray-cycle}, 
 $\epsilon$-Eri \citep{Coffaro_2020_eps_eri_xray_cycle}, 
 61~Cyg A \& B \citep{Hempelmann_2006_61Cygni_xraycycle, Robrade_2012_xray_61CygA_and_alpha_Cen_AB}, 
and Proxima~Cen \citep{Wargelin_2024}. Activity cycle variations are larger towards higher-energy, lower wavelengths compared to the visible wavelengths of the \ion{Ca}{ii} H\&K lines. In solar cycle 24 for example, the S-index of the \ion{Ca}{ii} H\&K activity indicators was $\sim$5.5 per cent larger during the phase of activity maximum compared to activity  minimum \citep{Egeland_2017_S_index_of_sun}, 
while the integrated flux emitted from the soft X-ray bin (see Section \ref{sec:XUV_cycle_Sun}) is a factor of $\sim$4 times larger, with even larger variations at shorter wavelengths. Of the current few sample of stars with X-ray cycles, only older stars with low ratios of $L_{\text{X-ray}}/L_{\text{bol}}$ have been found to exhibit an X-ray cycle similar to that of the Sun, with large, smooth amplitude modulations evolving over similar decadal timescales. Younger, faster rotating stars with moderate to high $L_{\text{X-ray}}/L_{\text{bol}}$ ratios instead exhibit faster, more complex X-ray cycles with smaller variations, if a cycle is apparent at all \citep{Ayres_2025_Landscape_Coronal_X-Ray_Var_and_Cycles}. \citet{Olah_2016_Magnetic_cycles_at_different_ages_of_stars} previously set an age of 2-3\,Gyr as a marker distinguishing between these two mentioned types of activity cycles. 

In order to model hydrodynamically escaping planetary atmospheres, a good constraint on the XUV flux received by the planet is essential. Unfortunately, absorption by the interstellar medium and Earth's atmosphere complicates this, limiting observational EUV constraints to either nearby systems \citep{France_2016, Yongblood_2016_MUSCLES} or empirical reconstructions reliant on X-ray observations \citep{Sanz-Forcada2011}. It is often necessary to adopt an XUV flux based off a proxy star when modelling the atmospheric of a specific planet. For these reasons, observing stellar activity cycles in X-ray can substantially improve our understanding of how cyclic XUV behaviour affects planetary atmospheric escape and its observables.

\label{sec:XUV_cycle_general}

\subsection{The XUV cycle of the Sun}
\label{sec:XUV_cycle_Sun}

\begin{figure}
    \includegraphics[width=0.475\textwidth]{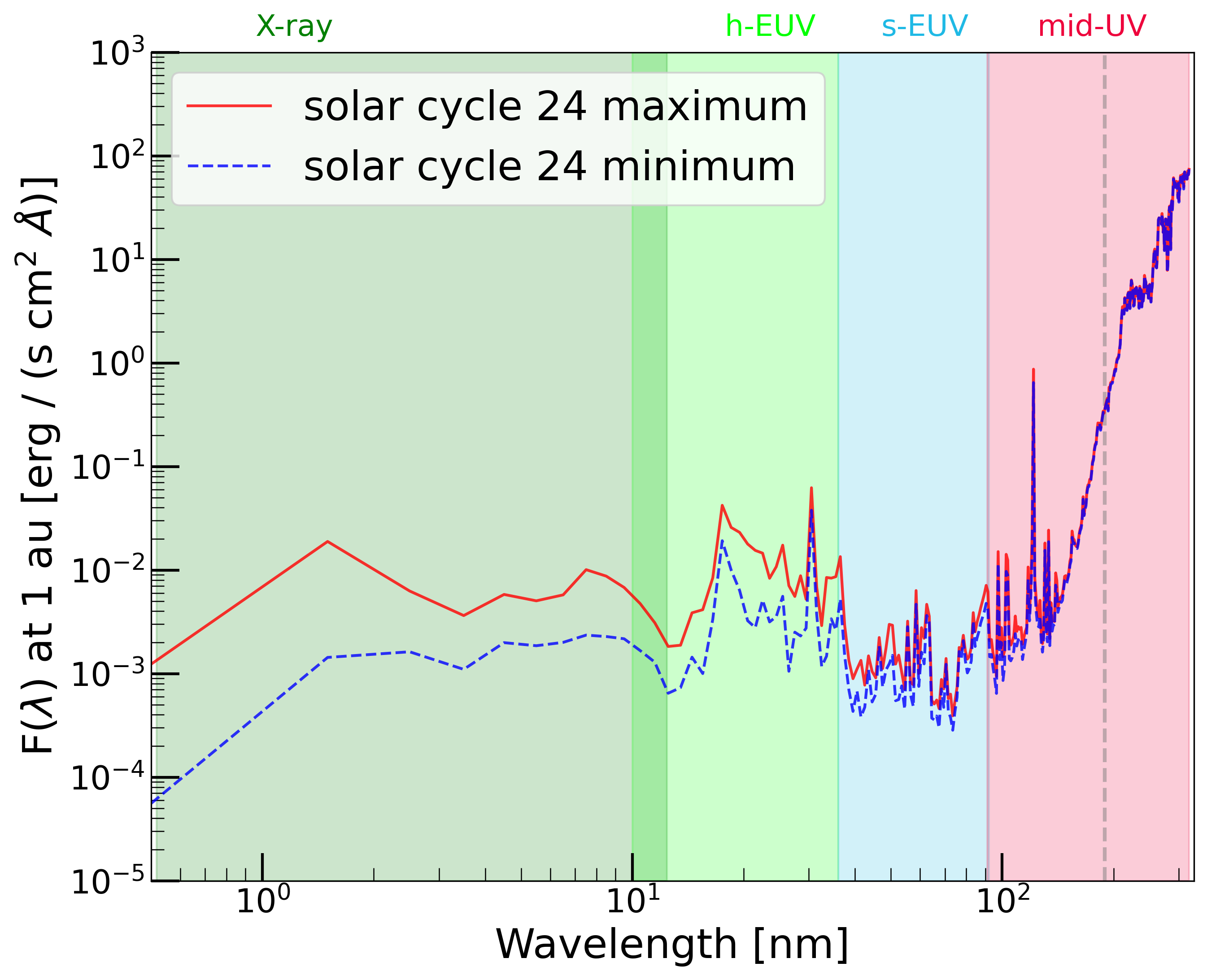}
     \includegraphics[width=0.475\textwidth]{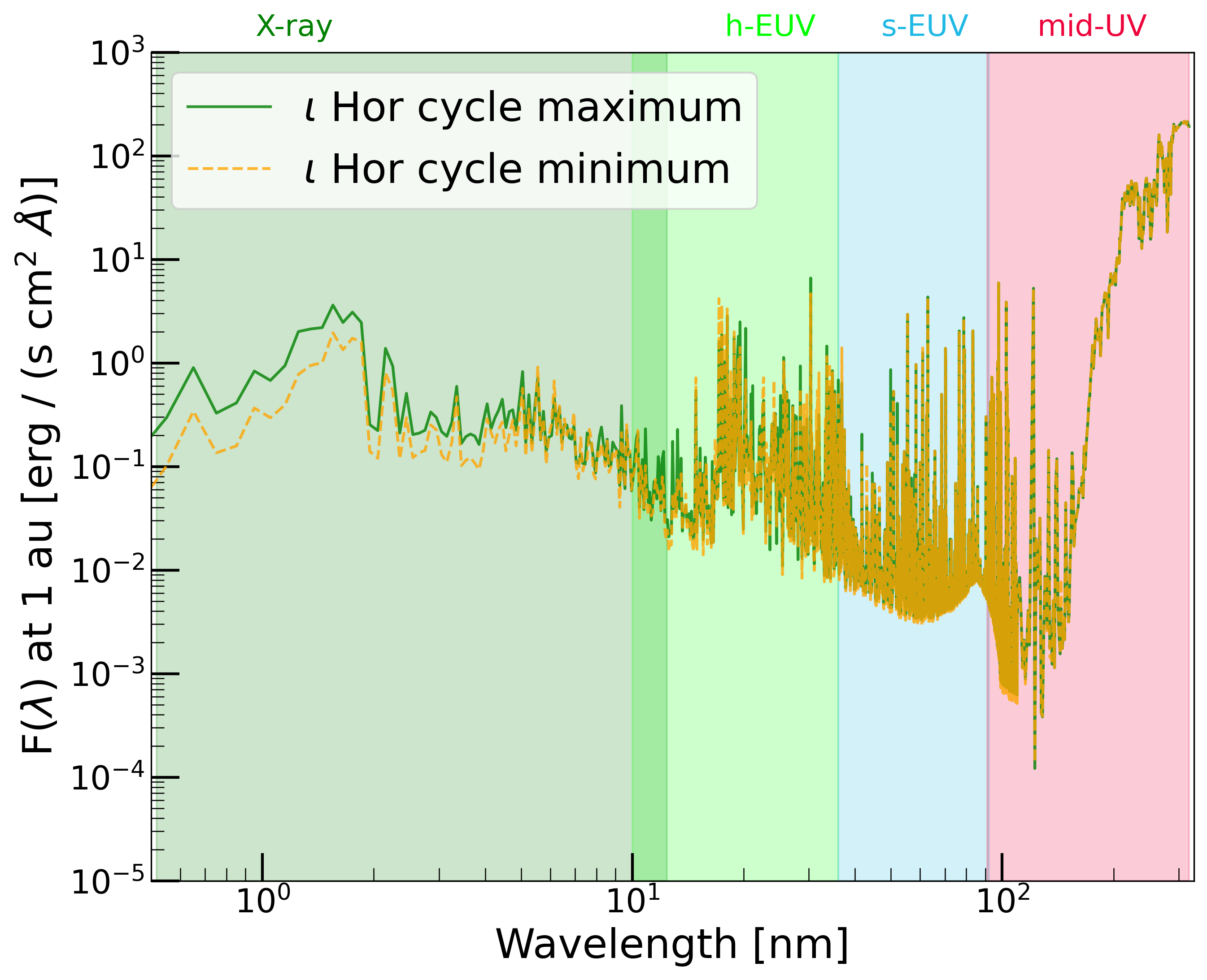}
    \caption{Upper panel: Solar SED at minimum and maximum phases of the solar activity cycle. The data at wavelengths below 189.5\,nm (marked by the vertical grey line) were obtained with the SEE instrument while above with the SORCE instrument as described in section \ref{sec:XUV_cycle_Sun}. Lower panel: Reconstructed SED of $\iota$~Hor during a minimum and maximum phase of the activity cycle, as described in section \ref{sec:SED_iHor}. The shaded regions common to both panels distinguish the wavelength bins used as input in our atmospheric escape model. From left to right they are X-ray (0.517-12.4\,nm), hard-EUV (10-36\,nm), soft-EUV (36-92\,nm) and mid-UV (91.2-320\,nm).
 }
 \label{fig:Sun_spec_min_max}
\end{figure} 

\begin{table}
\caption{Binned luminosities in the X-ray, hEUV, sEUV and mid-UV bands for the minimum and maximum phases of the activity cycle of the Sun (upper part of the Table) and of $\iota$~Hor (bottom part of the Table). 
In the final row, the ratio of the maximum to minimum luminosity for each wavelength band is given.}
\label{table:Star_lum_min_max}
\begin{tabular}{llllll}
\toprule
 & $L_{\text{X-ray}}$ & $L_{\text{hEUV}}$ & $L_{\text{sEUV}}$ & $L_{\text{mid-UV}}$ & EUV/mid-UV \\
 & ($L_{\odot}$) & ($L_{\odot}$) & ($L_{\odot}$) & ($L_{\odot}$) &  ratio \\ \midrule
\multicolumn{6}{|c|}{ \textbf{The Sun during cycle 24}} \\ \hline 
max & 4.98e-07 & 2.36e-06 & 8.68e-07 & 1.97e-02 & 1.64e-04 \\
min & 1.35e-07 & 9.24e-07 & 5.87e-07 & 1.94e-02 & 7.80e-05 \\
max/min & 3.69 & 2.55 & 1.48 & 1.02 & 2.10 \\
\hline 
\multicolumn{6}{|c|}{\textbf{$\iota$~Hor}} \\ \hline 
max & 3.33e-05 & 4.00e-05 & 2.81e-05 & 7.95e-02 & 8.56e-04 \\
min & 2.15e-05 & 3.98e-05 & 2.81e-05 & 7.95e-02 & 8.54e-04 \\
max/min & 1.55 & 1.00 & 1.00 & 1.00 & 1.00 \\
\bottomrule \end{tabular}
\end{table}

In order to obtain an XUV spectral energy distribution (hereby SED) during a minimum and maximum phase of the solar activity cycle, we can avail of numerous space-based spectral observations. In order to cover the wavelength range required by our atmospheric escape model, we combined datasets from two different observing missions. For wavelengths between 0.5-189.5\,nm 
we utilise spectral irradiance measurements from the Solar EUV Experiment (SEE) instrument of the NASA Thermosphere Ionosphere Mesosphere Energetics Dynamics (TIMED) mission \citep{Woods_2005_Solar_EUV_experiment}. For wavelengths between 189.5-320\,nm, we use measurements obtained by the SOlar Radiation and Climate Experiment (SORCE) NASA mission \citep{Woods_SORCE_overview_2021}. Of the SORCE dataset, the data we use between 189.5-310nm was obtained with the SOLar STellar Irradiance Comparison Experiment B (SOLSTICE-B) instrument while 310-320nm was obtained by the Spectral Irradiance Monitor (SIM) instrument. We select datasets from each mission representing daily averages. In the XUV wavelengths for which we use the SEE observations, we select the Level 3 dataset which is filtered to remove flares as to focus our study on longer timescale variations. The SEE data have coverage from the year 2002 to present while the SORCE data cover 2003 to 2020. From their combined time coverage, we select solar cycle 24 for our analysis. As pointed out by \citet{Hazra_2020}, the variation across this particular cycle is relatively weak compared to other cycles in terms of sunspot number variation \citep{Clette_2014_sunspot_number_400years}. In this sense, the resulting cyclic variation of planetary atmospheric escape and $\ion{He}{i}$~1083\,nm signature reported here for planets orbiting solar-like stars could be stronger, depending on the particular stellar cycle.

The upper panel of Figure \ref{fig:Sun_spec_min_max} displays the resulting SEDs during a minimum phase (dashed-blue) and maximum phase (solid-red) of the Sun's activity cycle. We select January 1 2009 and April 1 2014 to represent the minimum and maximum phases of the activity cycle. The  wavelength bins considered by our model set, distinguished by the background colour are soft X-ray (0.517–12.4 nm), hard-EUV (hEUV, 10–36 nm), soft-EUV (sEUV, 36–92 nm),
and mid-UV (91.2–320 nm), consistent with \citet{Allan_et_al_He_evol_2024, Allan_Vidotto_2025_young_He}. As previously mentioned, the activity cycle variation is greater towards shorter wavelengths. This is even more apparent in Figure \ref{fig:cycle_integ_fluxes} of Appendix \ref{appendix:XUV_cycle}, which displays the cyclic nature of the solar flux across each of our considered wavelength bins. 
The stronger cyclic variation at lower wavelengths is interesting in the context of observing planetary atmospheric escape using the helium triplet signature, as the higher energy XUV wavelengths which drive the escape vary substantially with the activity cycle, whereas the lower energy mid-UV photons capable of depopulating helium out of the 2$^3$s state via photoionisation exhibits negligible cyclic variation.
The values of the binned fluxes for the solar activity minimum and maximum used in our modelling is given in Table \ref{table:Star_lum_min_max}. 

\subsection{The XUV cycle of $\iota$~Hor}
 \label{sec:SED_iHor}

The star $\iota$-Hor is a young solar analogue of spectral type $\sim$ F8V/G0V \citep{Vauclair_2008_iota_hor} with a detected X-ray cycle \citep{Sanz_Forcada_2013_EARLIER_iota_hor_xray_cycle, Sanz_Forcada_2019_iota_hor_xray_cycle}. Compared to the Sun it has a much shorter cycle period of only 1.6 years \citep{Metcalfe_2010} with smaller amplitude variations. 

In order to study a stellar activity cycle differing from that of our Sun, we reconstruct the XUV SED of $\iota$~Hor during a minimum and maximum phase of its activity cycle. To achieve this, we utilise X-ray observations obtained by the XMM-Newton mission \citep{Turner_2001_XMM, Struder_2001_XMM} covering the star's coronal cycle from \citet{Sanz_Forcada_2019_iota_hor_xray_cycle}. We modelled the corona and transition region combining the information from UV lines from \citet{Sanz_Forcada_2019_iota_hor_xray_cycle} with the results of a 3-temperature fit to XMM-Newton EPIC spectra. We used the six highest observed fluxes in order to obtain the activity maximum SED in the X-ray, and conversely the lowest six fluxes for the activity minimum SED. The X-ray emission at the maximum was modelled with $\log T_{1,2,3}$(K)$= 6.18^{+0.06}_{-0.03}, \, 6.67^{+0.01}_{-0.01}, \, 6.96^{+0.01}_{-0.01}$ and $\log EM_{1,2,3}$(cm$^{-3}$)$=50.98^{+0.07}_{-0.09}, \, 51.23^{+0.02}_{-0.02}, \, 50.97^{+0.03}_{-0.03}$. At the minimum we got $\log T_{1,2,3}$(K)$= 6.10^{+0.03}_{-0.05}, \, 6.60^{+0.04}_{-0.02}, \, 6.87^{+0.03}_{-0.02}$ and $\log EM_{1,2,3}$(cm$^{-3}$)$=50.95^{+0.12}_{-0.07}, \, 50.97^{+0.07}_{-0.05}, \, 50.71^{+0.06}_{-0.15}$.
The coronal and transition region contribution to the SED was then modelled as in other stars with no high-resolution X-rays data \citep[see][]{Sanz-Forcada_2025_HeI_10830}.

For the photospheric contribution, we availed of synthetic modelling by \citet{Castelli_Kurucz_2003_new_grid_ATLAS9} adopting the model with T=6250\,K, $\log g = 4.5$ and solar abundance. We find that this provides a good match to the HST/STIS spectra ranging 2600-2800 \AA\ reported in \citet{ama23}. We used the same model for maximum and minimum of the cycle.
A stellar radius of 1.185 \(R_\odot\) was assumed in the mentioned $\iota$~Hor SED calculations, as well as in our ray-tracing model for predicting the $\ion{He}{i}$~1083\,nm planetary transit signature. In modelling the planetary atmospheric escape of these $\iota$~Hor-orbiting planets, a stellar mass of 1.16 \(M_\odot\) was assumed.

The lower panel of Figure \ref{fig:Sun_spec_min_max} displays the resulting activity minimum and maximum SEDs of $\iota$~Hor, while Table \ref{table:Star_lum_min_max} lists the luminosities in the wavelength bins relevant to our modelling. It is clear that the variation between our produced minimum and maximum activity SEDs are minor and limited to the shorter wavelengths. This is unsurprising considering the smaller amplitude variation of the fast $\iota$~Hor activity cycle compared to that of the Sun.

\section{Modelling planetary atmospheric escape and the $\ion{He}{i}$~1083\,\protect\lowercase{nm} transit signature}

\label{sec:modelling_planets}

\begin{table}
\caption{Planetary inputs assumed for each considered planet classification.}
\label{Tab:escape_inputs}
\begin{tabular}{lllll}
\toprule
 & sub-Neptune & Neptune-mass & Saturn-mass & HJ \\ \midrule
M$_{ \text{pl} }$ (M$_{\text{Nep}} )$ & 0.41 & 1.45 & 5.62 & 12.64 \\
R$_{\text{pl} }$ (R$_{\text{Nep}} )$ & 0.55 & 1.04 & 2.34 & 3.92 \\
$R_{\text{base}}$ (R$_{\text{pl} }$) & 2.07 & 1.57 & 1.32 & 1.23 \\
$g_{\text{surf}}$ (m s$^{-2}$) & 14.98 & 14.92 & 11.45 & 9.15 \\
\bottomrule \end{tabular}
\end{table}

In modelling the planetary atmospheric escape and associated $\ion{He}{i}$~1083\,nm signatures presented in this work, we utilised the model previously presented in \citet{Allan_Vidotto_2025_young_He}. This one dimensional model numerically solves the hydrodynamic atmospheric escape self-consistently, while simultaneously determining the population of helium in its 2$^3$S state as required for producing the $\ion{He}{i}$~1083\,nm signature. As described in \citet[][section 2.2]{Allan_Vidotto_2025_young_He}, we again utilise the lower atmosphere model of \citet{Parmentier_Guillot_I} in selecting the lower boundary conditions, setting the lower boundary of our hydrodynamic escape model for the upper atmosphere to occur at a pressure of 10\,nbar. For each planet, we assume a constant He/H number density fraction of 0.1. 
In predicting the $\ion{He}{i}$~1083\,nm signatures, we utilise the ray-tracing model used in \citet{Allan_Vidotto_2025_young_He} and references therein. We assume a circular orbit and a transit along the centre of the stellar disk (impact parameter $b=0$) resulting in a transit duration of  $ t_{\rm dur}  \approx \frac{2 R_\star}{\sqrt{G M_\star/a}} $, where $G$ is the gravitational constant and $a$ is the planetary orbital distance. Depending on the stellar SED used, the stellar mass and radius, $M_\star$ and $R_\star$ is set to either that of the Sun or $\iota$~Hor. 

In order to study how a stellar activity cycle affects the planetary atmospheric escape and the $\ion{He}{i}$~1083\,nm signature of a variety of exoplanets, we consider four different types of planets at two different orbital distances, 0.025\,au and 0.05\,au. We consider theoretical sub-Neptune (sub-Nep), Neptune-mass (Nep-mass), Saturn-mass (Sat-mass) and hot-Jupiter (HJ) exoplanets. Table \ref{Tab:escape_inputs} displays the assumed planetary masses and radii and the corresponding surface gravities. The assumed masses and radii were inspired by that of the sub-Neptune TOI-1430b \citep{Zhang_4_mini_Nep}, the Neptune-massed GJ436b \citep{von_Braun_2012_pl_params}, Saturn-massed TOI-1268b \citep{Subjak_2022_TOI1268b, Perez_Gonzales_2023_TOI-1268b} and hot-Jupiter HD209458b \citep{Bonomo_2017A&A...602A.107B}, all of which have had attempted $\ion{He}{i}$~1083\,nm transmission spectroscopy.

\section{Findings of our study}
\label{sec:results}

\subsection{The effect of a solar-like activity cycle on the atmospheric escape of highly irradiated exoplanets}
\label{sec:solar-like_cycle_hydro_effects}

\begin{figure}
    \includegraphics[width=0.48\textwidth]{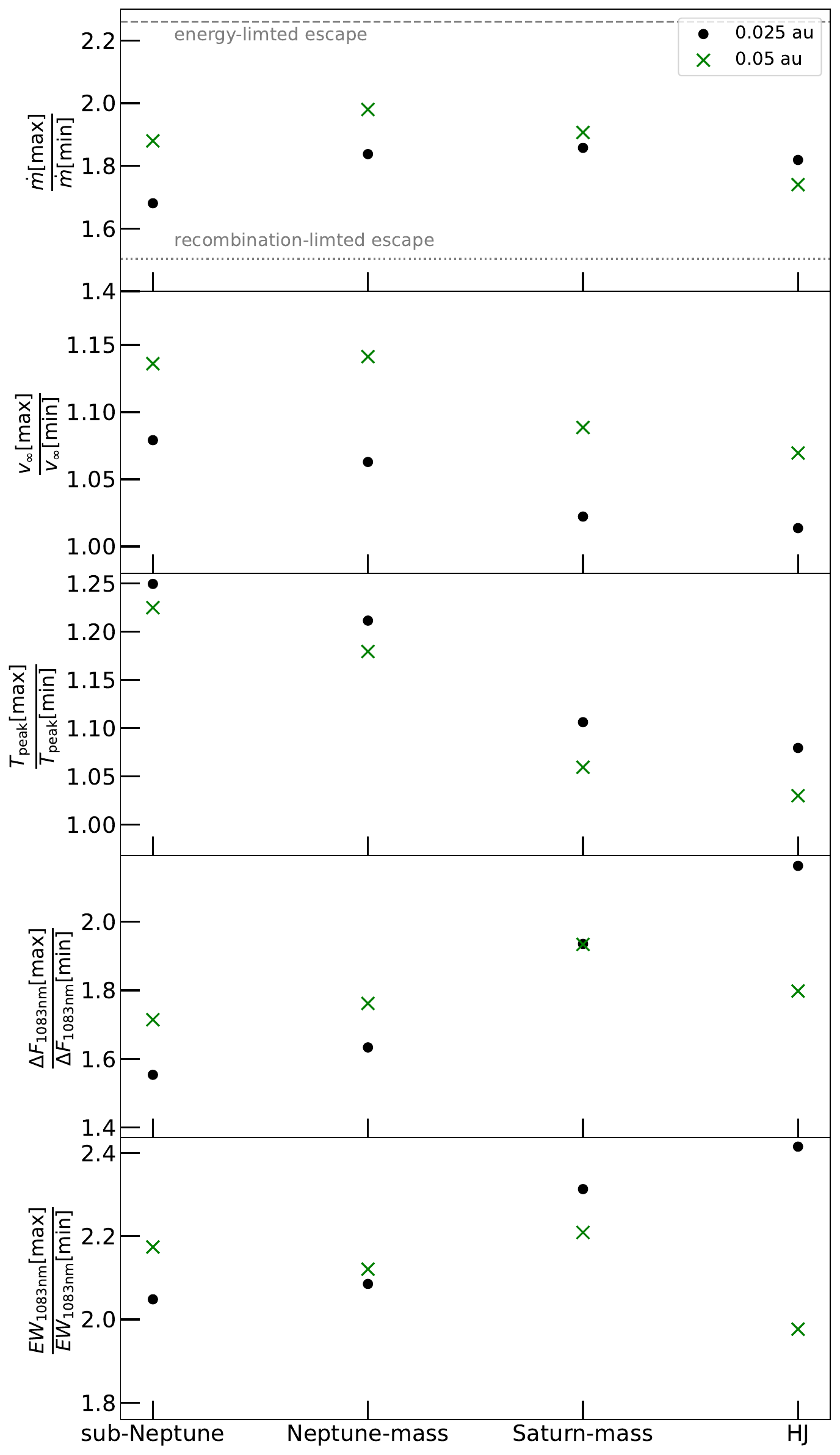} 
    \caption{
        Maximum-to-minimum cycle variation in various properties of the escaping atmosphere for inner 0.025\,au (circles) and outer 0.05\,au (crosses) orbits around the Sun-like star. The absolute values of each parameter are listed in Table \ref{Tab:escape_predictions_solar_all}. The horizontal  lines in the upper panel mark expected mass-loss rate variations based on the change in EUV flux for two different regimes of atmospheric escape as explained in the text. 
         }
 \label{fig:increase_params}
\end{figure}

In this section, we discuss how the solar-like activity cycle previously described in section \ref{sec:XUV_cycle_Sun} affects the planetary atmospheric escape predictions.
Table \ref{Tab:escape_predictions_solar_all} 
lists the predicted hydrodynamic escape properties of each modelled planet during either a maximum or minimum phase of the solar activity cycle, as indicated by the square brackets. 
The upper three panels of Figure \ref{fig:increase_params} displays the relative variations in these hydrodynamic properties at a maximum phase of the activity cycle compared to a minimum, for the same orbital distance. 
Clearly, a solar-like activity cycle affects the atmospheric escape of the various types of highly irradiated exoplanets considered.
For each of the modelled planets, the variation in mass-loss rate due to the activity cycle phase is substantial, with the smallest increase from the maximum to minimum phase being a factor of 1.68 for the sub-Neptune planet at 0.025\,au, and the highest being a factor of ~2 for the Neptune-mass planet at 0.05 au. While the terminal velocities are also influenced by the stellar activity cycle, their variations are considerably less than those of the mass-loss rates, with variations of only 1--2 km/s. The stronger dependence of the activity cycle with the mass-loss rates compared to the terminal velocities is due to the photoionisation heating occurring predominantly within the sub-sonic rather than the super-sonic region of each of the modelled planetary atmospheres.

It is also apparent from Figure \ref{fig:increase_params} that the activity cycle has a greater influence on the atmospheric escape of the more distant planets. This is reflected in the larger variations in mass-loss rate at the further orbital distance of 0.05\,au (green cross) compared to that at 0.025\,au (black circle), for all but the HJ planet. 
This stronger activity cycle influence at the further orbital distance is likely due to a transition in the efficiency of atmospheric escape. Going from higher to lower irradiation levels, atmospheric mass-loss rate is known to transition from the less efficient recombination-limited regime within which $\dot{m} \propto \sqrt{F_{\text{XUV}}}$ to the more efficient energy-limited escape regime where $\dot{m} \propto F_{\text{XUV}}$ \citep{Murray-Clay2009, Owen_Alvarez_2016, Caldiroli_2022_evap_eff}. 
For the solar cycle 24 we consider in this work, the fraction of high-energy EUV flux\footnote{We find in our model that only the flux included in our hEUV and sEUV wavelength bins contribute to atmospheric heating and escape, hence for this discussion we consider only the activity cycle variation in the `EUV' combined wavelength range of these two bins, omitting contributions from our X-ray bin.} received by each planet during the activity maximum relative to activity minimum is $2.26$. Hence, a corresponding mass-loss rate variation factor of 2.26 between activity cycle maximum and minimum would be indicative of energy-limited escape, as marked by the dashed-line in the upper panel of Figure \ref{fig:increase_params}. While a factor of $\sqrt{2.26}=1.50$ would indicate recombination-limited escape, marked by the dotted-line. 
The predicted mass-loss rates exhibit activity cycle variation factors between 1.68-1.86 for the planets at 0.025\,au and 1.74-1.98 for those at 0.05\,au. The modelled atmospheric escape is hence between the energy-limited and recombination-limited regimes, with the closer orbiting planets being pushed closer to recombination-limited escape.

\begin{table}
\caption{Hydrodynamic escape predictions for planets orbiting a star with a solar-like activity cycle. For each property, the predicted value obtained during the maximum activity cycle is first given, followed by that for the minimum activity phase. $\dot{m}$ refers to the planetary mass-loss rate, $v_{\infty}$ the wind terminal velocity (velocity at the outermost edge of the modelled atmosphere) and $T_{\text{peak}}$ to the peak of the non-isothermal atmospheric temperature profile.
}
\label{Tab:escape_predictions_solar_all}
\begin{tabular}{lllll}
\toprule
 & sub-Nep & Nep-mass & Sat-mass & HJ \\ 
 \hline 
\multicolumn{5}{|c|}{\textbf{0.025\,au}} \\ \hline
$\dot{m}[\text{max}]$~g/s & 8.8e+12 & 2.7e+12 & 1.2e+12 & 7.8e+11 \\
$\dot{m}[\text{min}]$~g/s & 5.2e+12 & 1.5e+12 & 6.3e+11 & 4.3e+11 \\
$v_{\infty}[\text{max}]$~km/s & 68 & 67 & 64 & 61 \\
$v_{\infty}[\text{min}]$~km/s & 67 & 66 & 63 & 60 \\
$T_{\text{peak}}[\text{max}]$~K & 8782 & 8848 & 9349 & 9391 \\
$T_{\text{peak}}[\text{min}]$~K & 7028 & 7303 & 8452 & 8700 \\
\hline 
\multicolumn{5}{|c|}{\textbf{0.05\,au}} \\ \hline
$\dot{m}[\text{max}]$~g/s & 2.1e+12 & 6.4e+11 & 2.5e+11 & 1.1e+11 \\
$\dot{m}[\text{min}]$~g/s & 1.1e+12 & 3.2e+11 & 1.3e+11 & 6.2e+10 \\
$v_{\infty}[\text{max}]$~km/s & 32 & 31 & 27 & 23 \\
$v_{\infty}[\text{min}]$~km/s & 30 & 28 & 25 & 22 \\
$T_{\text{peak}}[\text{max}]$~K & 5674 & 6064 & 8571 & 9134 \\
$T_{\text{peak}}[\text{min}]$~K & 4632 & 5142 & 8090 & 8869 \\

\bottomrule \end{tabular}
\end{table}

\subsection{The effects of a solar-like activity cycle on the \ion{He}{i}~1083\,nm transit signature of highly irradiated exoplanets} 
\label{sec:effects_solar_cycle_helium_sig}

\begin{figure*}
\centering
  {\Large \textbf{The Sun's XUV cycle effects on $\ion{He}{i}$~1083\,nm transit signature predictions}}\\[0.5em]  
    \begin{subfigure}{0.95\textwidth}
        \includegraphics[width=0.95\textwidth]{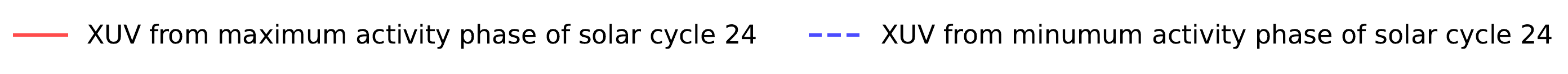}
  \end{subfigure}
  \hfill
  \begin{subfigure}{0.48\textwidth}
        \includegraphics[width=0.95\textwidth]{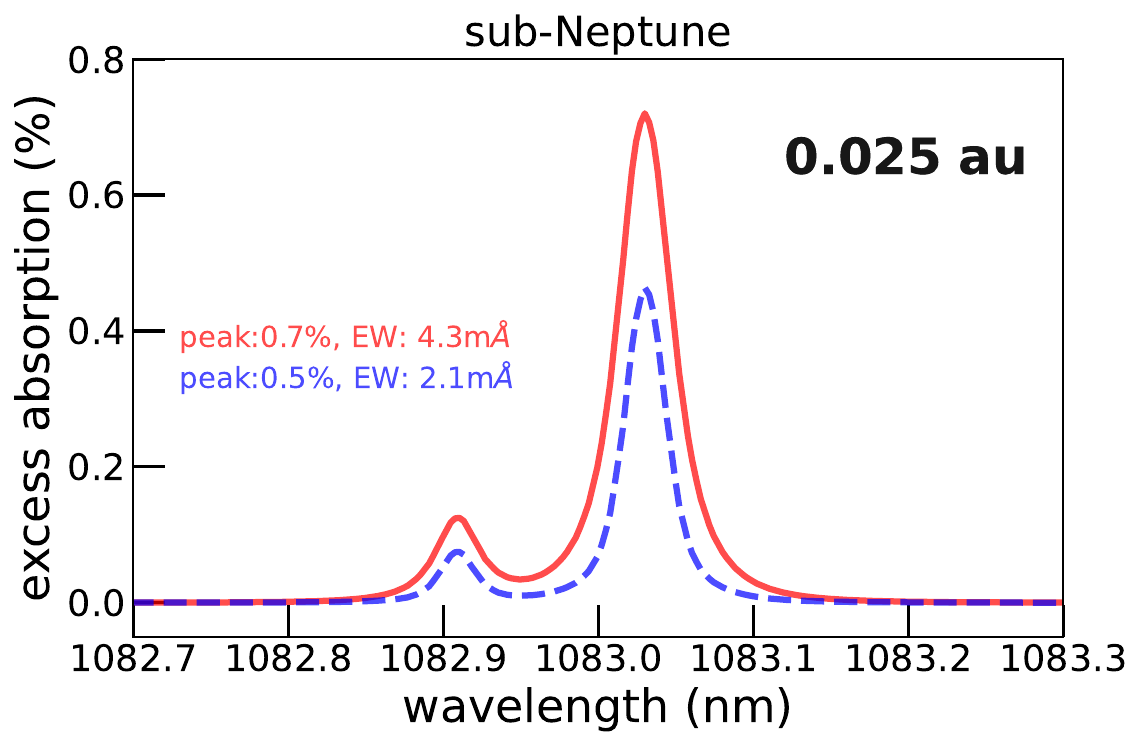}
  \end{subfigure}
  \hfill
  \begin{subfigure}{0.48\textwidth}
        \includegraphics[width=0.95\textwidth]{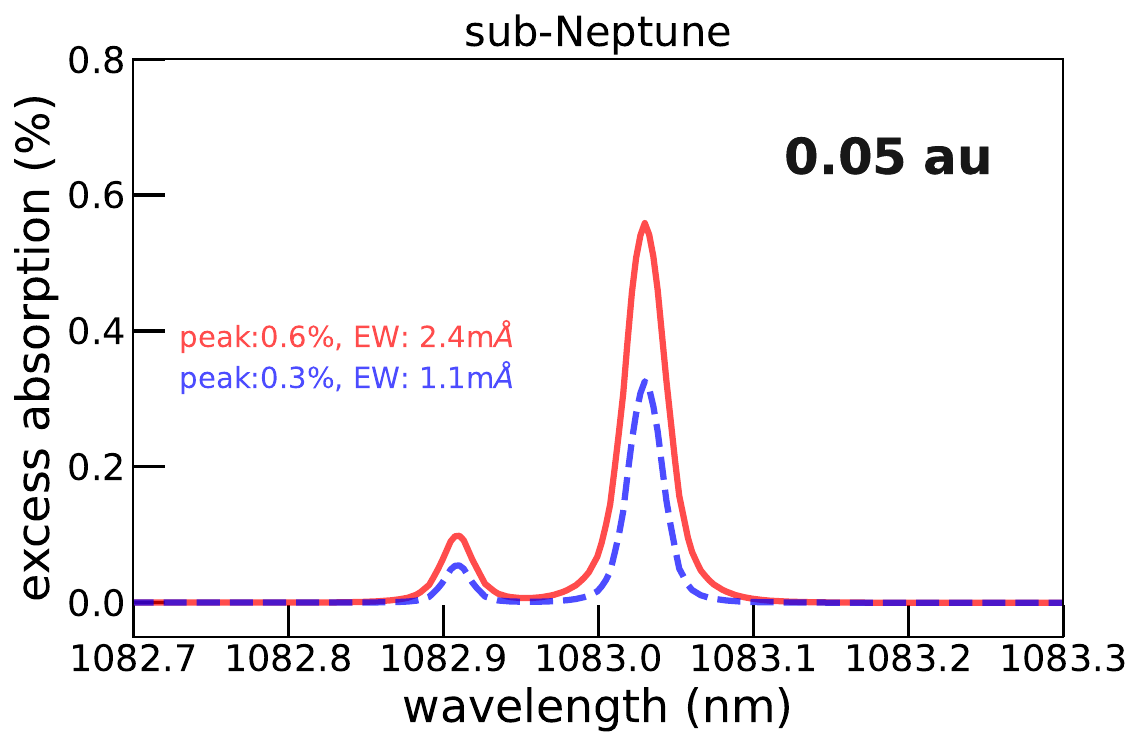} 
  \end{subfigure}
  \begin{subfigure}{0.48\textwidth}
        \includegraphics[width=0.95\textwidth]{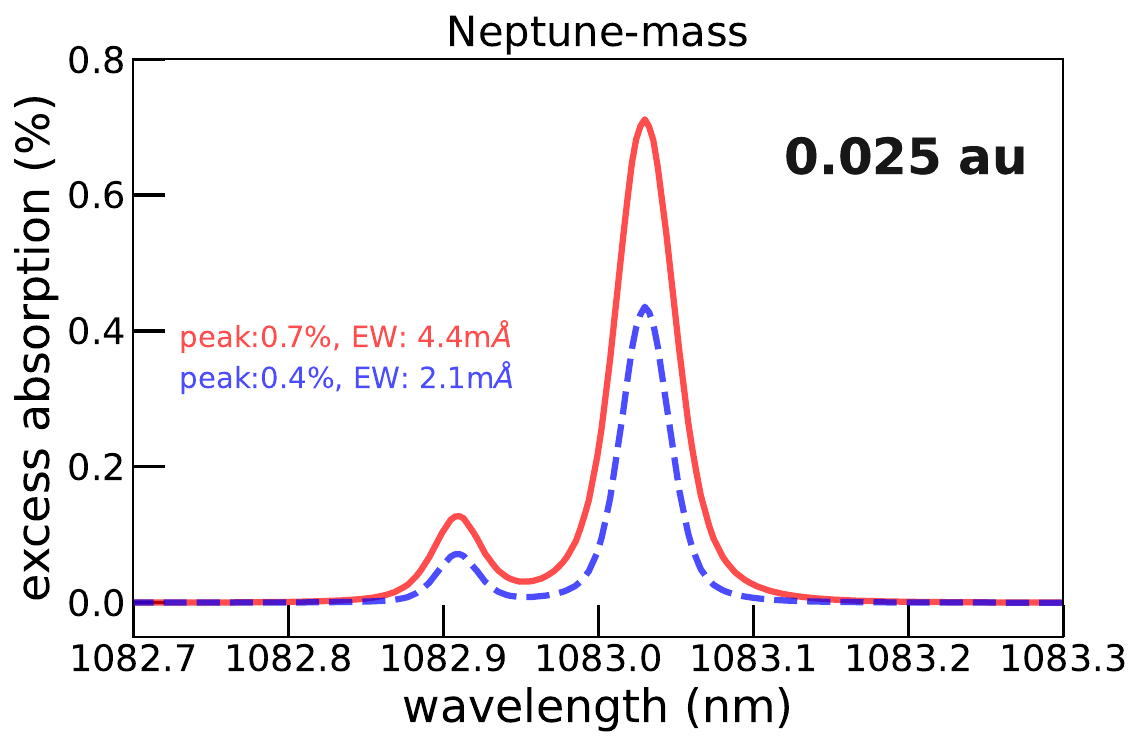} 
  \end{subfigure}
  \hfill
  \begin{subfigure}{0.48\textwidth}
        \includegraphics[width=0.95\textwidth]{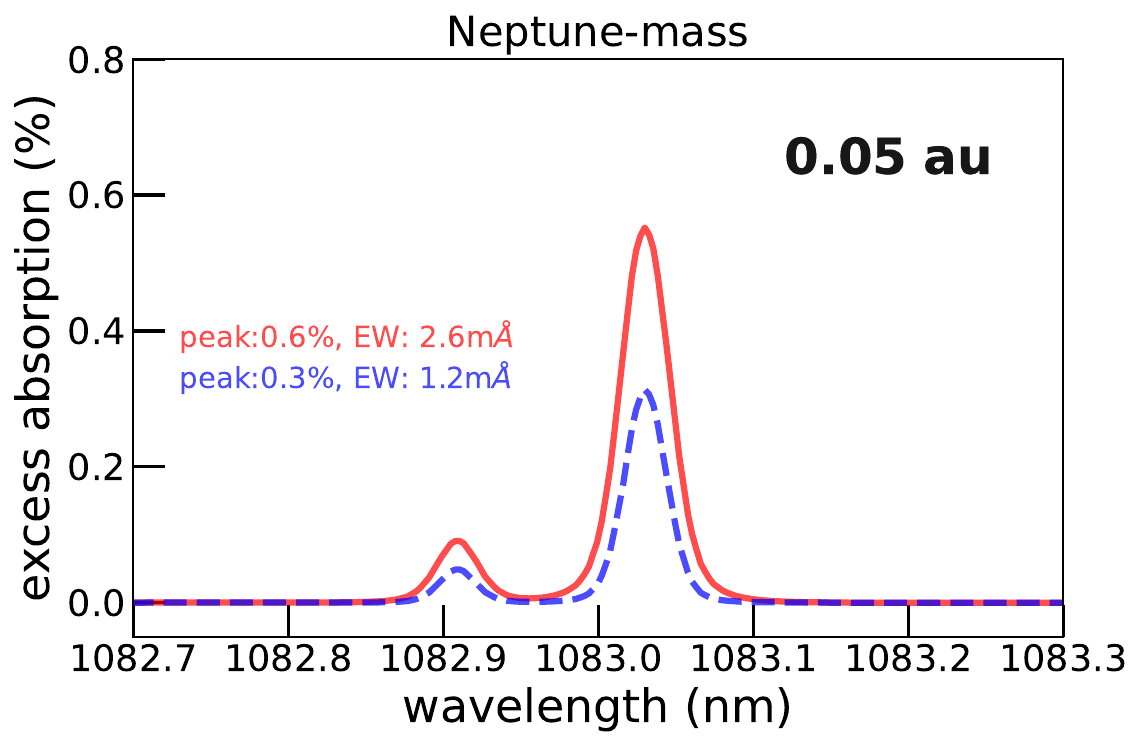}  
  \end{subfigure}

  \begin{subfigure}{0.48\textwidth}
        \includegraphics[width=0.95\textwidth]{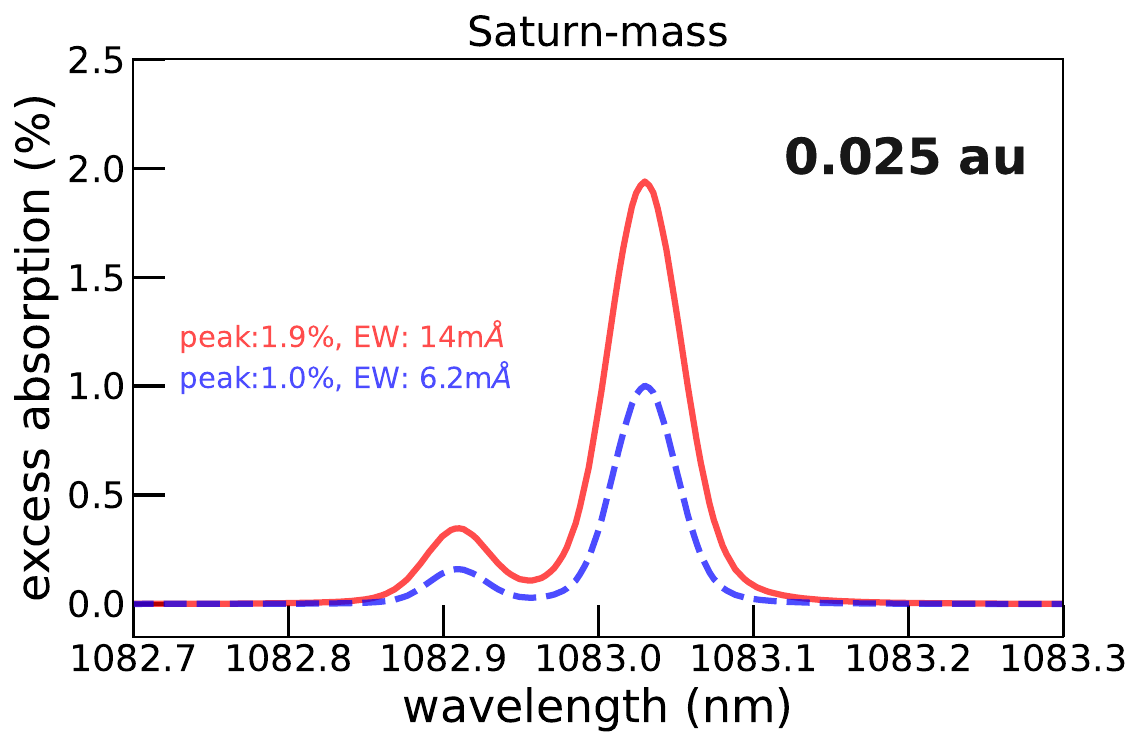} 
  \end{subfigure}
  \hfill
  \begin{subfigure}{0.48\textwidth}
        \includegraphics[width=0.95\textwidth]{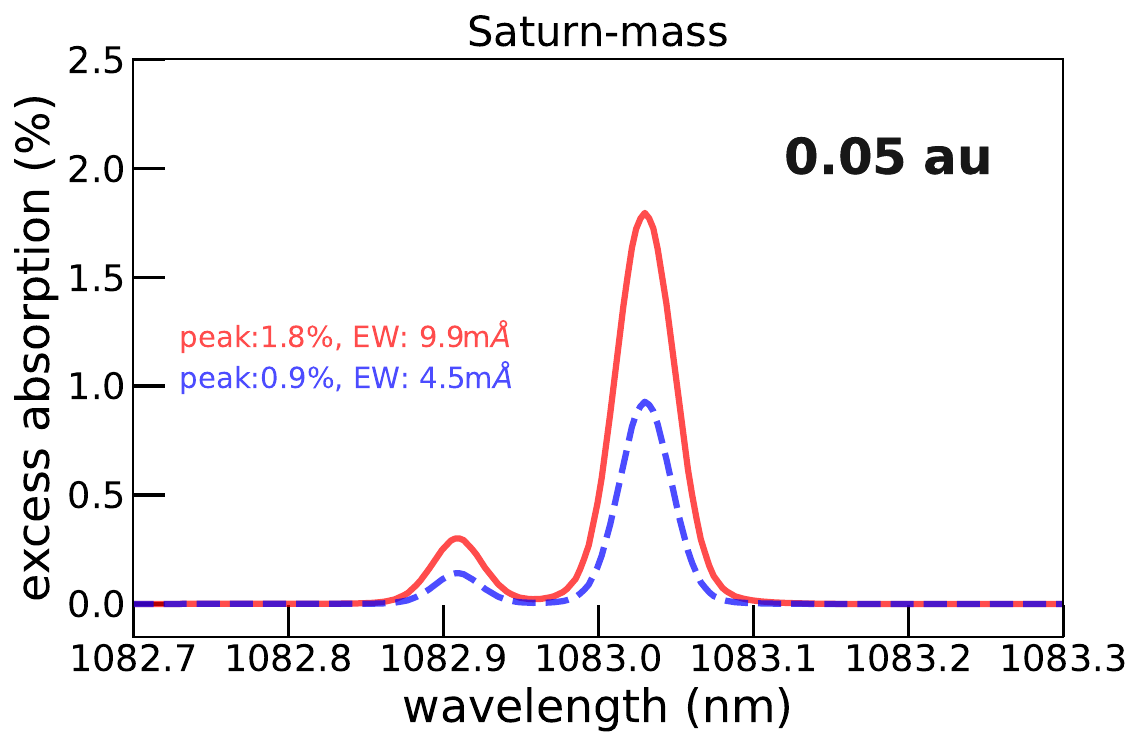} 
  \end{subfigure}
  \begin{subfigure}{0.48\textwidth}
        \includegraphics[width=0.95\textwidth]{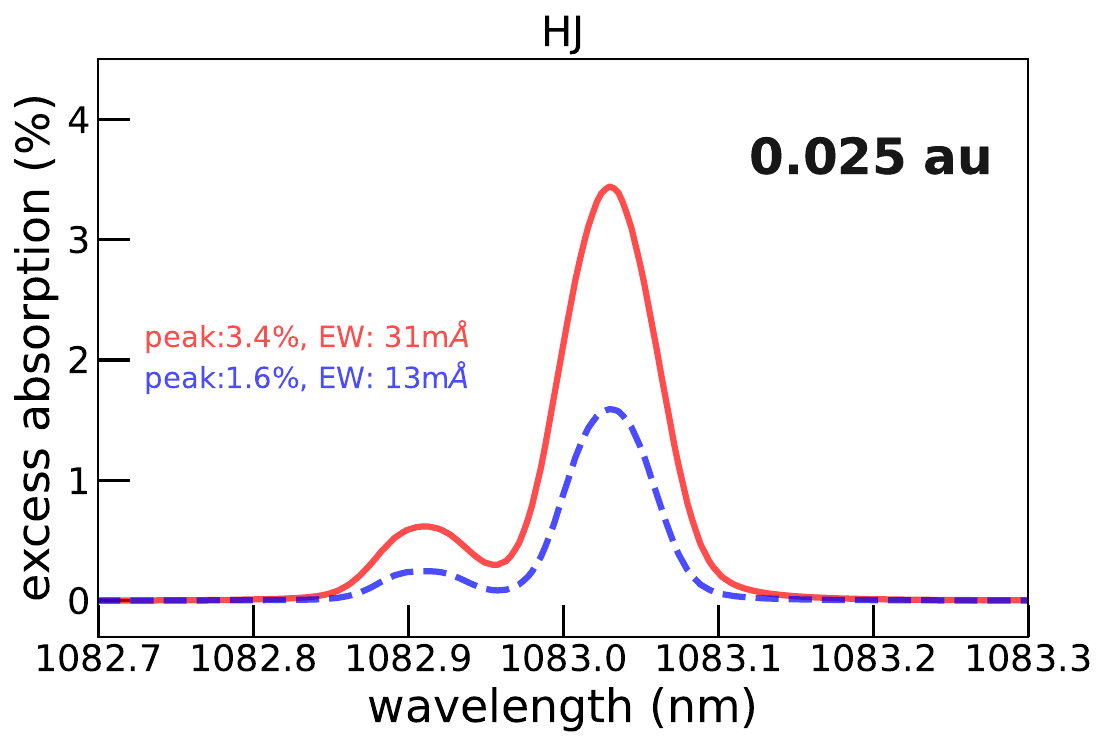}  
  \end{subfigure}
  \hfill
  \begin{subfigure}{0.48\textwidth}
        \includegraphics[width=0.95\textwidth]{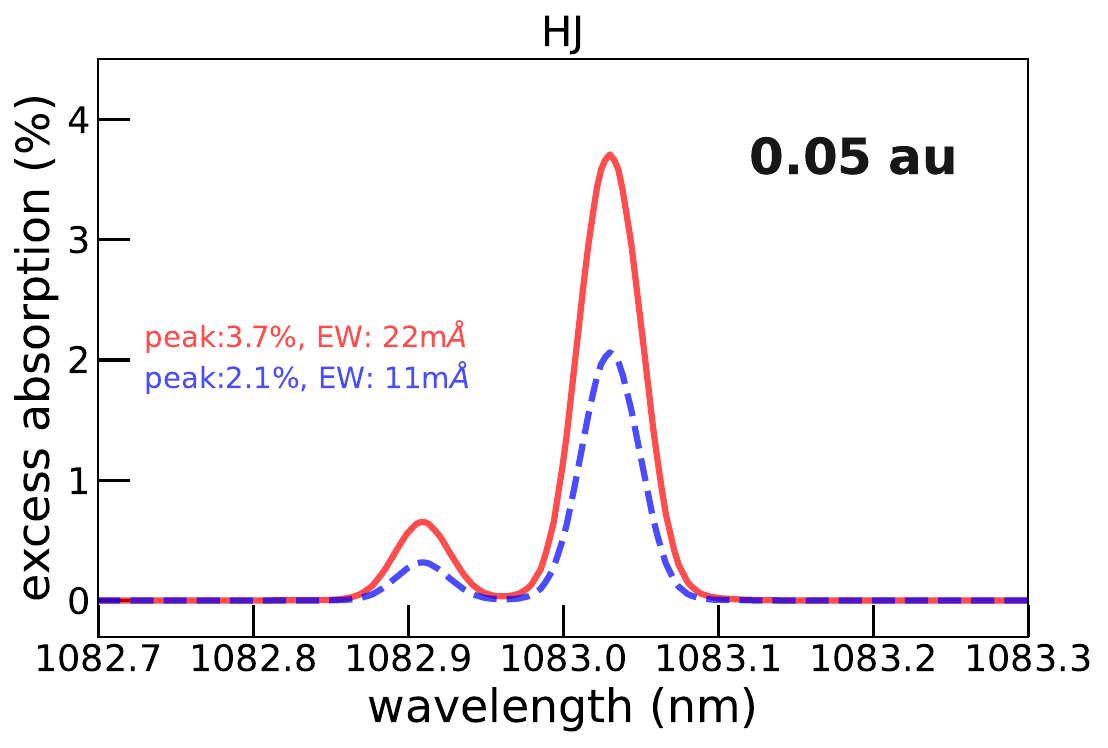}  
  \end{subfigure}

  \caption{Predicted transit-phase-averaged helium triplet profiles assuming high energy fluxes consistent with maximum (red) and minimum (blue) stages during the solar activity cycle. The different rows correspond to the four different types of planets while the columns distinguish the assumed orbital distance. Note the differing y-scales for the differing planet types.}
  \label{fig:triplet}
\end{figure*}

It was shown in the previous subsection that a solar-like activity cycle can significantly affect the atmospheric escape of highly irradiated exoplanets. We now focus on the extent to which such atmospheric escape variations affect the planet's observable $\ion{He}{i}$~1083\,nm transit signatures. Figure \ref{fig:triplet} displays the predicted transit-phase-averaged $\ion{He}{i}$~1083\,nm profiles for each modelled planet if observed during a maximum phase (red, solid line) or minimum phase (blue, dashed line) of a solar-like activity cycle. The profile's peak excess absorption and equivalent width (EW) are listed in each panel. The factor by which these two properties vary from cycle maximum to minimum are shown in the lower two panels of Figure \ref{fig:increase_params}, respectively.
Evidently, a solar-like activity cycle can significantly influence the $\ion{He}{i}$~1083\,nm transit signature of highly irradiated exoplanets. The stronger and faster atmospheric escape at a phase closer to the activity cycle maximum produces deeper, broader $\ion{He}{i}$~1083\,nm absorption features.  
Activity cycle variations of the Sun's emitted mid-UV flux, capable of depopulating helium via photoionisation out of its metastable $2^3$S state are insignificant relative to the larger variation in XUV fluxes, responsible for driving atmospheric escape, as shown previously in Figure \ref{fig:Sun_spec_min_max} and Table \ref{table:Star_lum_min_max}. Hence, enhancement of the $\ion{He}{i}$~1083\,nm transit feature at activity cycle maximum by stronger atmospheric escape far outweighs the marginal $\ion{He}{i}$~1083\,nm weakening effect of slightly more mid-UV photons available to depopulate the observationally important \ion{He}{I}~($2^3$S) state.

\begin{figure}
\begin{subfigure}{0.5\textwidth}
        \includegraphics[width=0.95\textwidth]{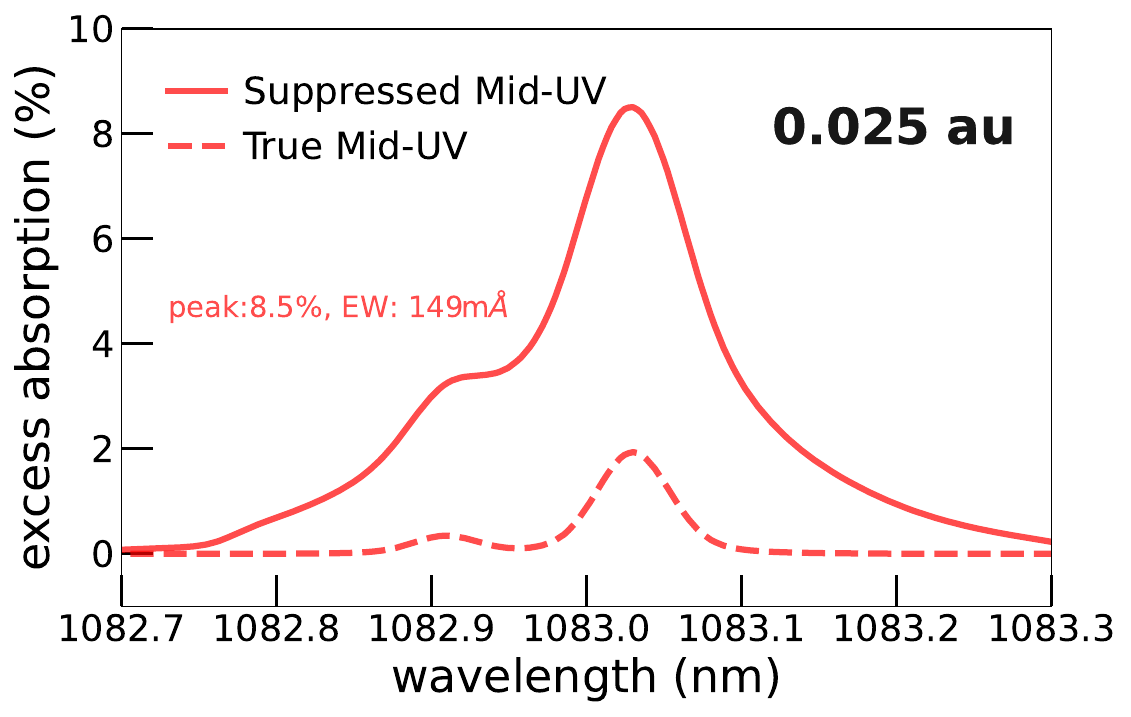}
  \end{subfigure}
  \hfill
  \begin{subfigure}{0.5\textwidth}
    \includegraphics[width=0.95\textwidth]{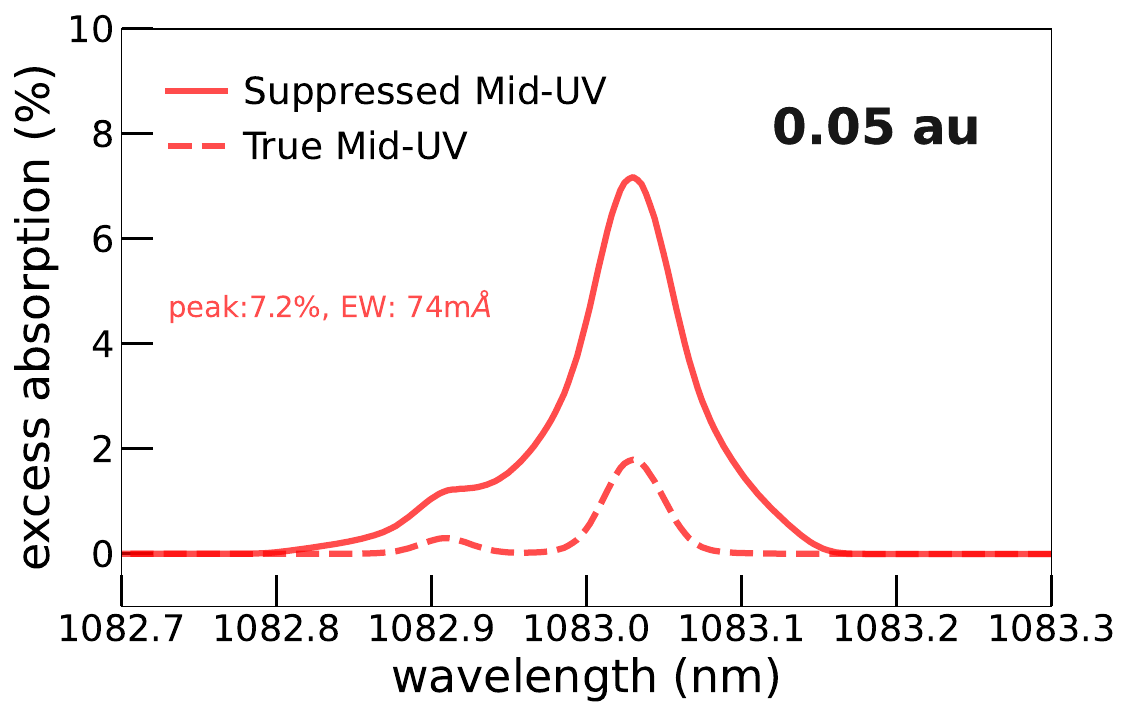} 
  \end{subfigure}
    \caption{ 
        Demonstrative test suppressing the mid-UV flux received by the modelled Saturn-mass planet orbiting a solar-like star during a maximum phase of its activity cycle. For the mid-UV suppressed case (solid-line) the mid-UV flux capable of depopulating helium out of the $2^3$S state has been reduced by a factor of 1000. The upper and lower panels correspond to orbital distances of 0.025 and 0.05\,au, respectively. For comparison, the smaller $\ion{He}{i}$~1083\,nm profile of the original modelled signature with the true mid-UV flux is shown by the dashed profile. 
         }
 \label{fig:triplet_mid_UV_supressed_Sun}
\end{figure}

Despite atmospheric escape being naturally weaker at larger orbital distances (Table \ref{Tab:escape_predictions_solar_all}), the $\ion{He}{i}$~1083\,nm transit signature and popular `tracer of atmospheric escape' interestingly weakens only slightly for planets orbiting further from a solar-like star (Figure \ref{fig:triplet}), even strengthening for some of the modelled planets at further orbits of the more active $\iota$~Hor (as we will see in Figure \ref{fig:triplet_IotaHor}). A similar non-monotonic behaviour of $\ion{He}{i}$~1083\,nm absorption with orbital distance was previously found by \citet{Biassoni_2024}, using the ATmospheric EScape (\texttt{ATES}) model \citep{Caldiroli_2021_ATES}. \citet{Biassoni_2024} proposed that larger extensions of the modelled atmosphere, which they set to the Hill radius $R_{\text{Hill}} \approx a \big( \frac{M_\text{pl}}{3M_\star} \big)^{1/3} $, is responsible for the cases of stronger $\ion{He}{i}$~1083\,nm absorption at further orbital distances. However, the modelled atmospheric extension of the model used here does not depend on the orbital distance. As mentioned in \citet{Allan_Vidotto_2025_young_He}, the outer atmospheric extension is set to the minimum distance required for the modelled atmospheric grid to fully encompasses the stellar disk during mid-transit, $R_\star$ for a transit impact parameter of $b=0$ as is assumed for the modelled transits here. Hence, an alternative explanation for the $\ion{He}{i}$~1083\,nm behaviour with orbital distance we find is required. We propose that the weaker mid-UV flux received by the planets at further orbital distances of the solar-like star is instead responsible. Mid-UV photons are capable of reducing the $\ion{He}{i}$~1083\,nm signature by photoionising helium out of its observationally important $\ion{He}{i}$($2^3$S) state \citep{Oklopcic_2019_dep_st_rad}. The weaker mid-UV flux received by the further orbiting planets can cause the helium in their weaker escaping atmospheres to be less effectively depopulated out of its $2^3$S state. 
To show this, Figure \ref{fig:triplet_mid_UV_supressed_Sun} displays the results of a demonstrative test case in which the mid-UV fluxes received by the Saturn-mass planets orbiting around the solar-like star during activity maximum is artificially suppressed by a factor of 1000 (solid profiles) compared to that found using $L_{\text{mid-UV}}$ of Table \ref{table:Star_lum_min_max} (dashed profiles). Naturally, the test cases with suppressed mid-UV fluxes yield significantly stronger $\ion{He}{i}$~1083\,nm profiles compared to the profiles obtained using the true mid-UV fluxes (dashed profiles), despite their identical hydrodynamic escape predictions. This is because the lower number of mid-UV photons received leads to a reduction in photoionisations out of the $\ion{He}{i}$($2^3$S) state. Notably however, the mid-UV suppressed $\ion{He}{i}$~1083\,nm signature at 0.05\,au is weaker than that at the closer orbital distance of 0.025\,au. Hence, if the emitted solar flux at mid-UV wavelengths was significantly lower, the modelled planet's $\ion{He}{i}$~1083\,nm transit signature would better trace their atmospheric escape, being stronger at closer orbits where their escape is stronger. However, the $\ion{He}{i}$~1083\,nm behaviour with orbital distance is complex and is affected by more environmental properties than the mid-UV flux alone, as shown later in figure \ref{fig:triplet_mid_UV_supressed_Iota}. A perhaps counter-intuitive result of the behaviour is that observational studies seeking to detect $\ion{He}{i}$~1083\,nm transit signatures may benefit from selecting targets orbiting further from stars with strong mid-UV fluxes, despite their weaker atmospheric escape.

\subsection{Atmospheric escape with $\iota$~Hor as a stellar host}
\label{sec:results_iota_hor}

\begin{figure*}
\centering
  {\Large \textbf{ \ion{He}{i}~1083\,nm transit predictions with $\iota$~Hor as the host star}}\\[0.5em]  
  \begin{subfigure}{0.48\textwidth}
        \includegraphics[width=0.95\textwidth]{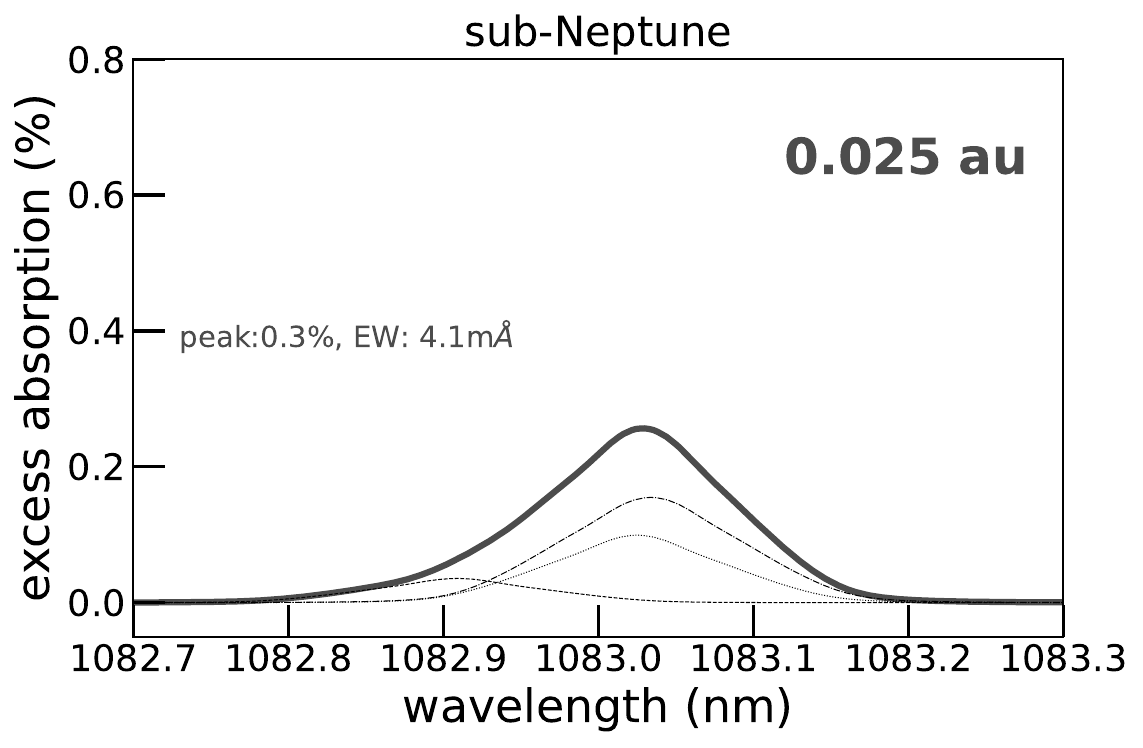}
  \end{subfigure}
  \hfill
  \begin{subfigure}{0.48\textwidth}
        \includegraphics[width=0.95\textwidth]{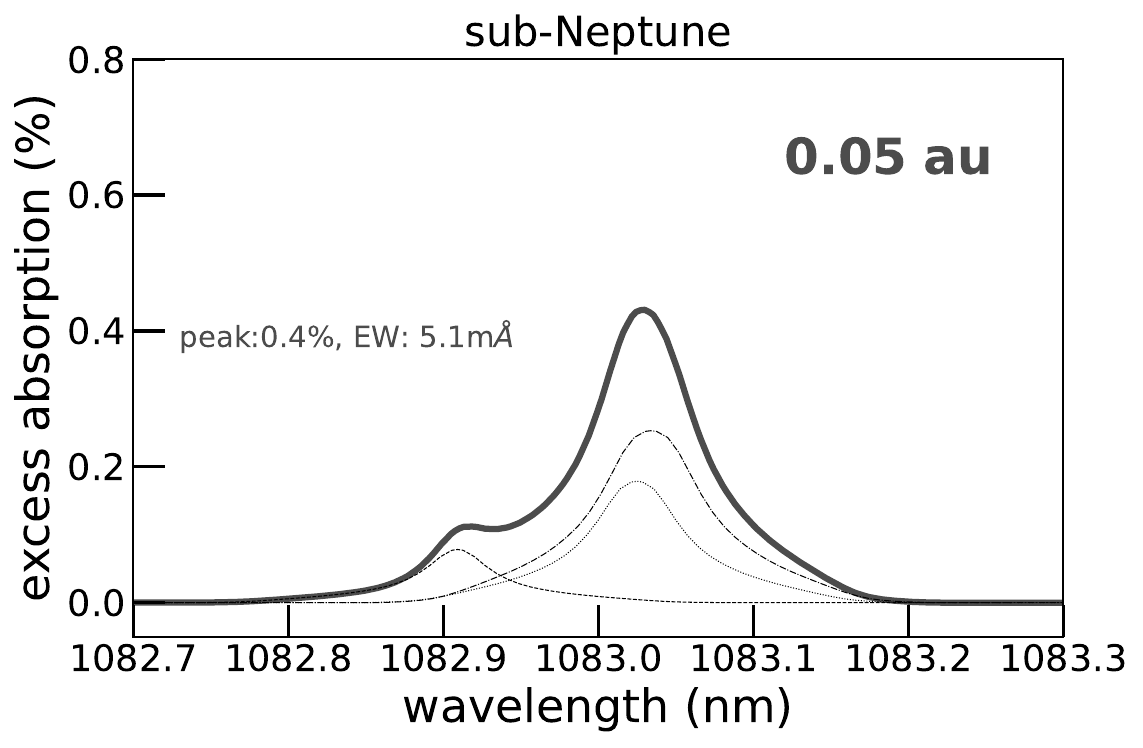} 
  \end{subfigure}
  \begin{subfigure}{0.48\textwidth}
        \includegraphics[width=0.95\textwidth]{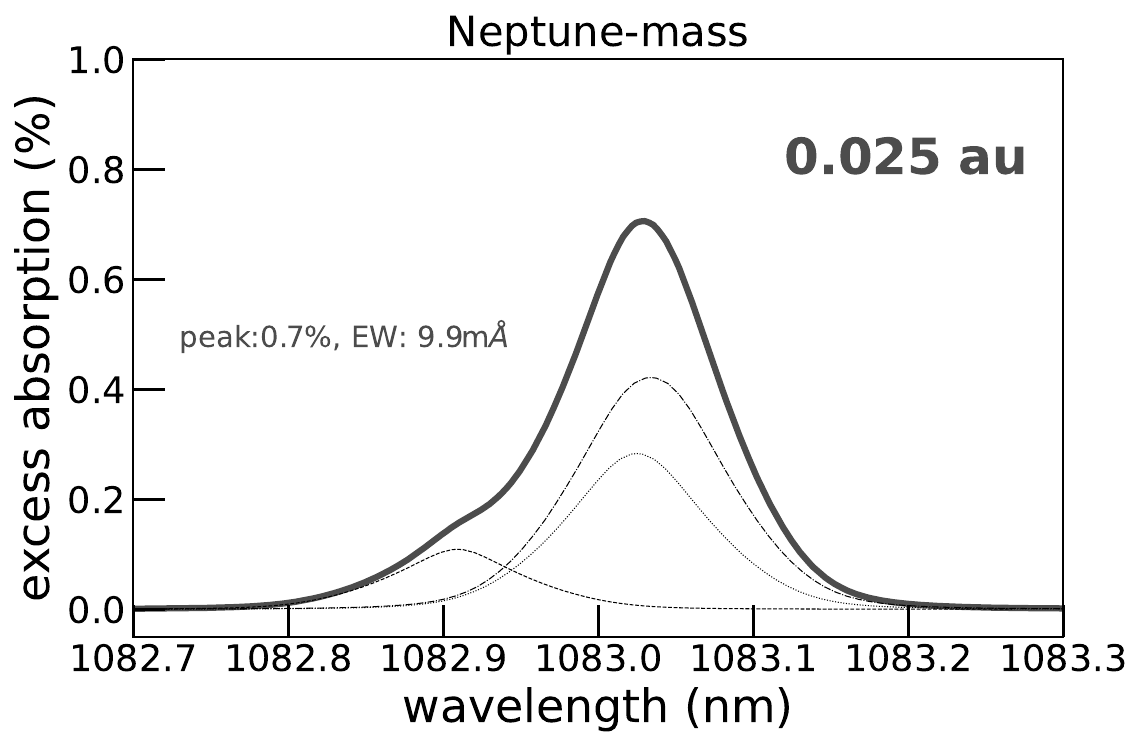} 
  \end{subfigure}
  \hfill
  \begin{subfigure}{0.48\textwidth}
        \includegraphics[width=0.95\textwidth]{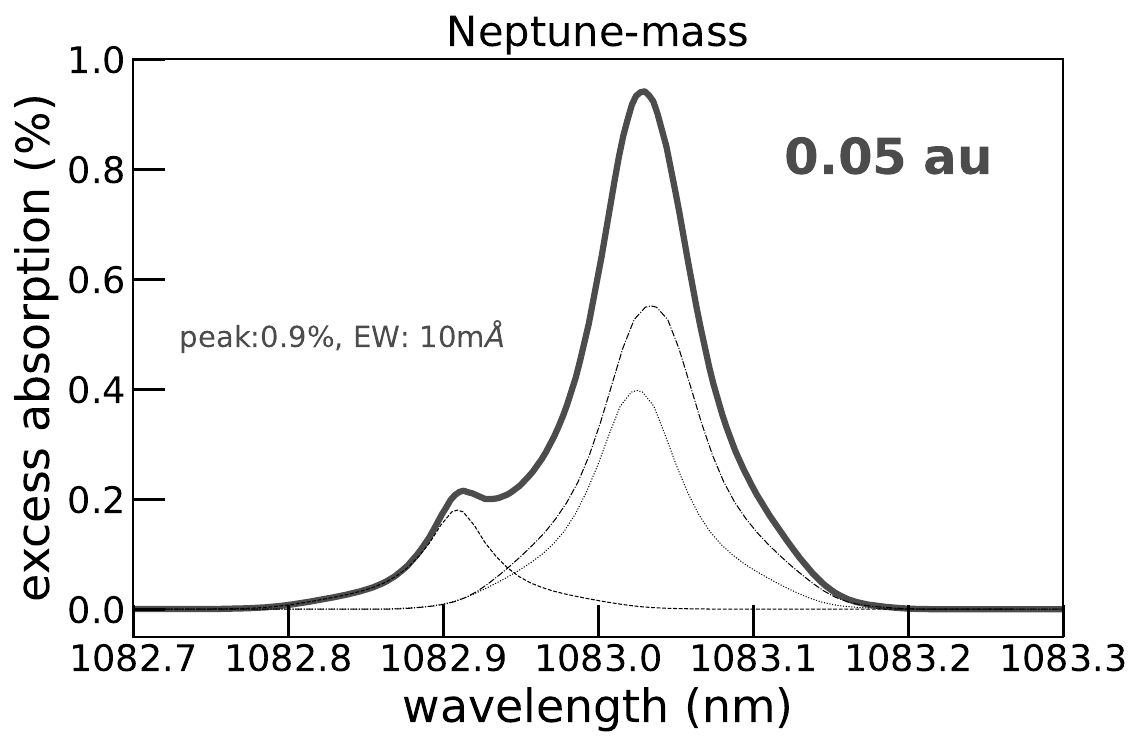}  
  \end{subfigure}

  \begin{subfigure}{0.48\textwidth}
        \includegraphics[width=0.95\textwidth]{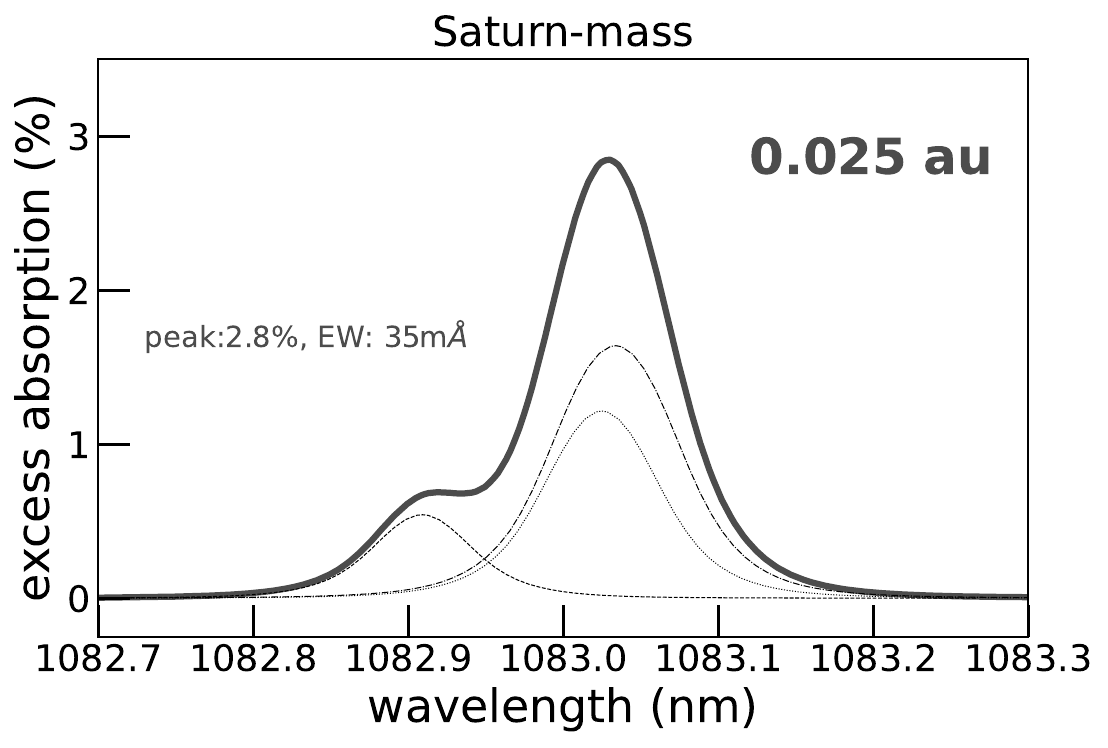} 
  \end{subfigure}
  \hfill
  \begin{subfigure}{0.48\textwidth}
        \includegraphics[width=0.95\textwidth]{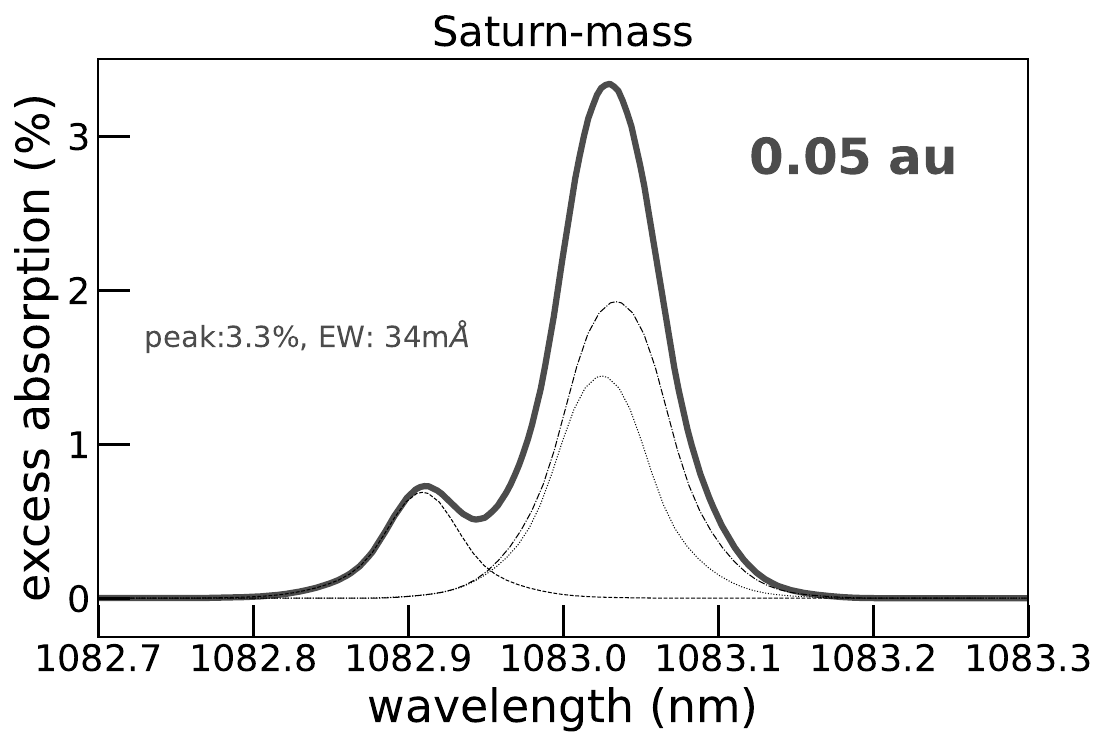} 
  \end{subfigure}
  \begin{subfigure}{0.48\textwidth}
        \includegraphics[width=0.95\textwidth]{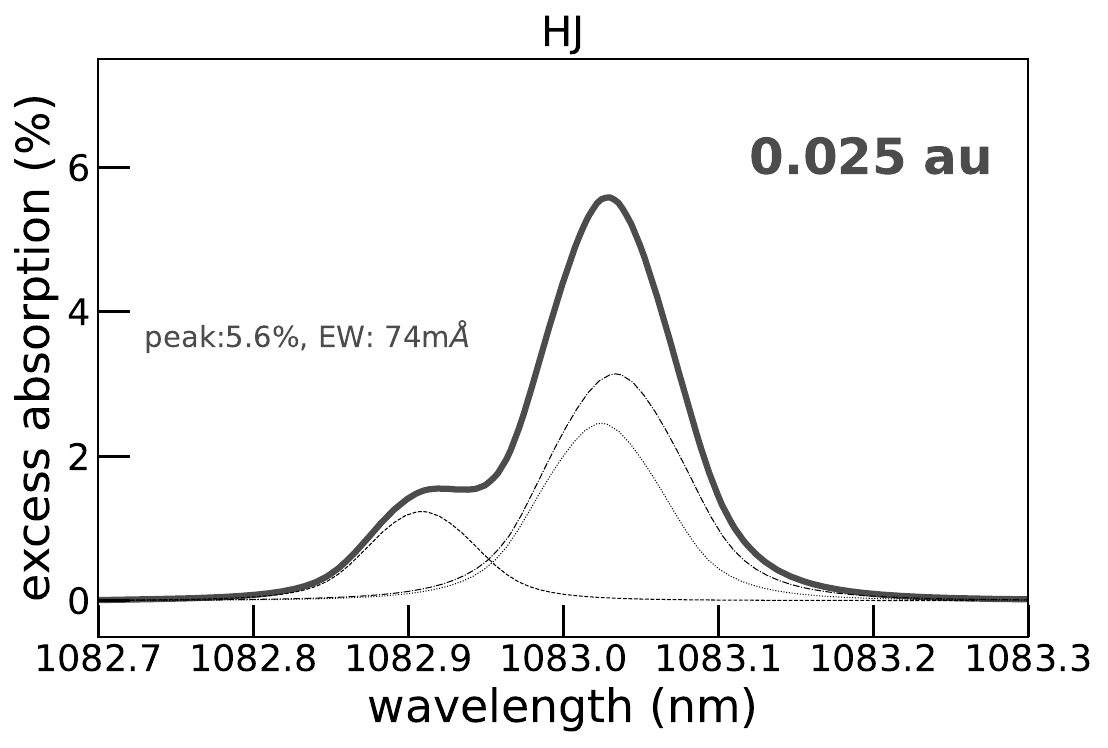}  
  \end{subfigure}
  \hfill
  \begin{subfigure}{0.48\textwidth}
     \includegraphics[width=0.95\textwidth]{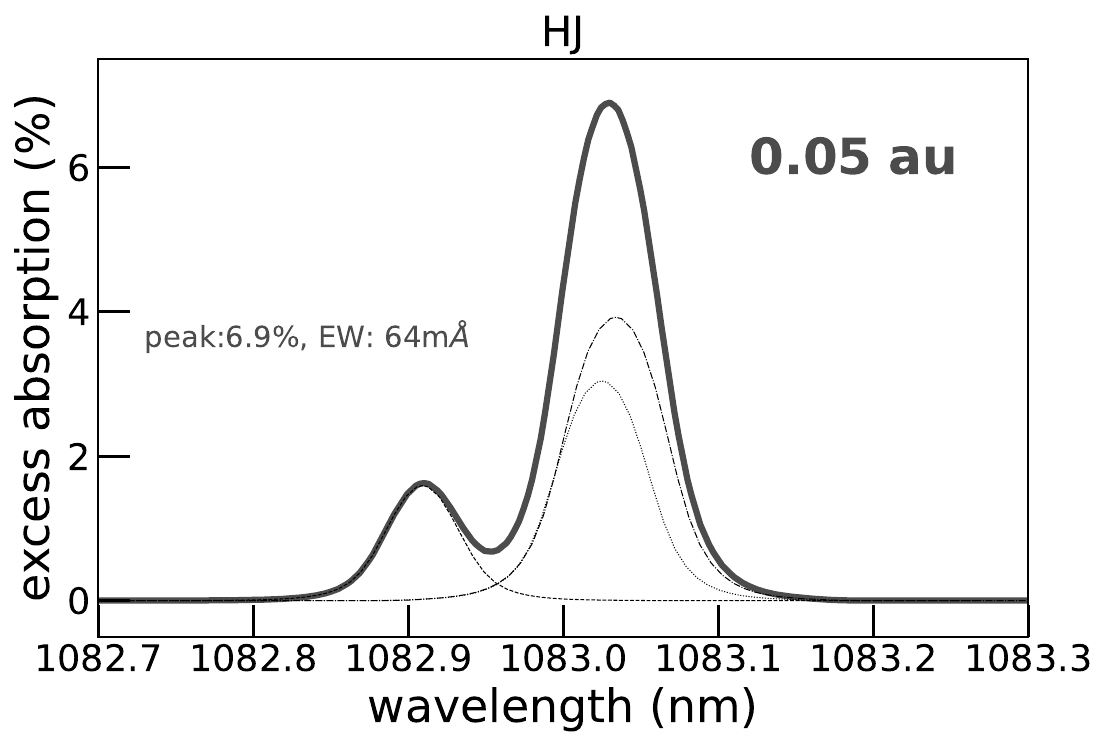} 
  \end{subfigure}

    \caption{
        Predicted helium triplet profiles assuming high energy fluxes obtained from the SED of $\iota$~Hor previously presented in section \ref{sec:SED_iHor}. The three lighter, non-solid lines distinguish individual contributions from the three lines comprising the triplet.  
         }
 \label{fig:triplet_IotaHor}
\end{figure*}

\begin{figure}
\begin{subfigure}{0.5\textwidth}
        \includegraphics[width=0.95\textwidth]{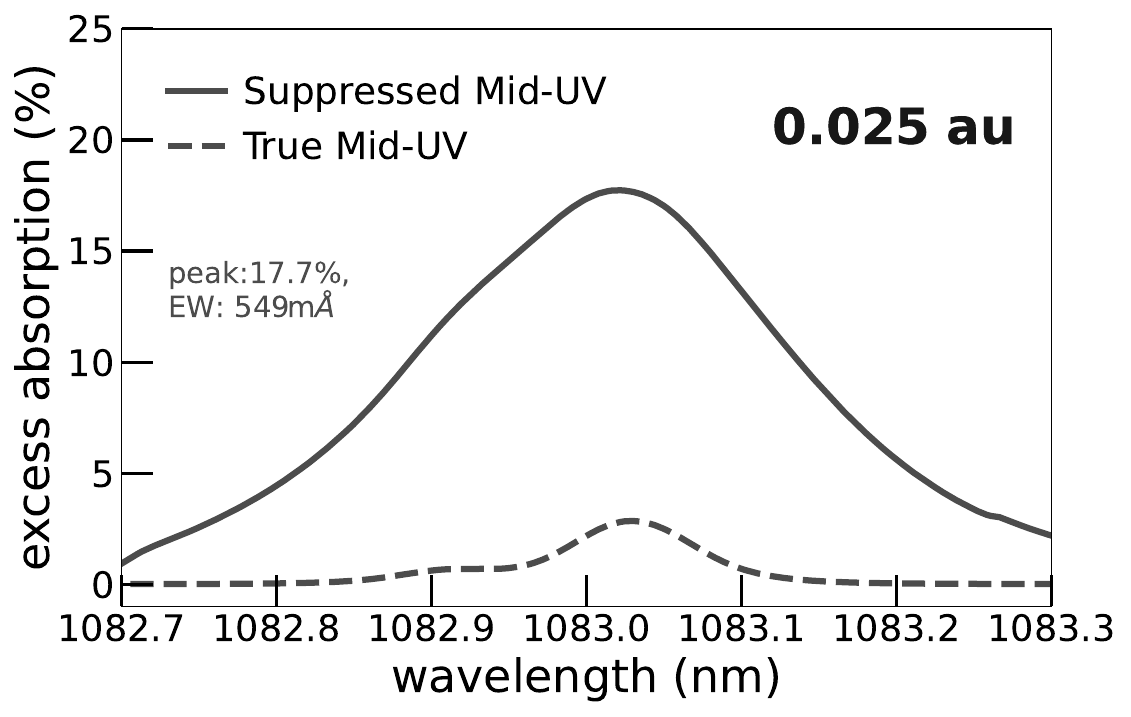}
  \end{subfigure}
  \hfill
  \begin{subfigure}{0.5\textwidth}
    \includegraphics[width=0.95\textwidth]{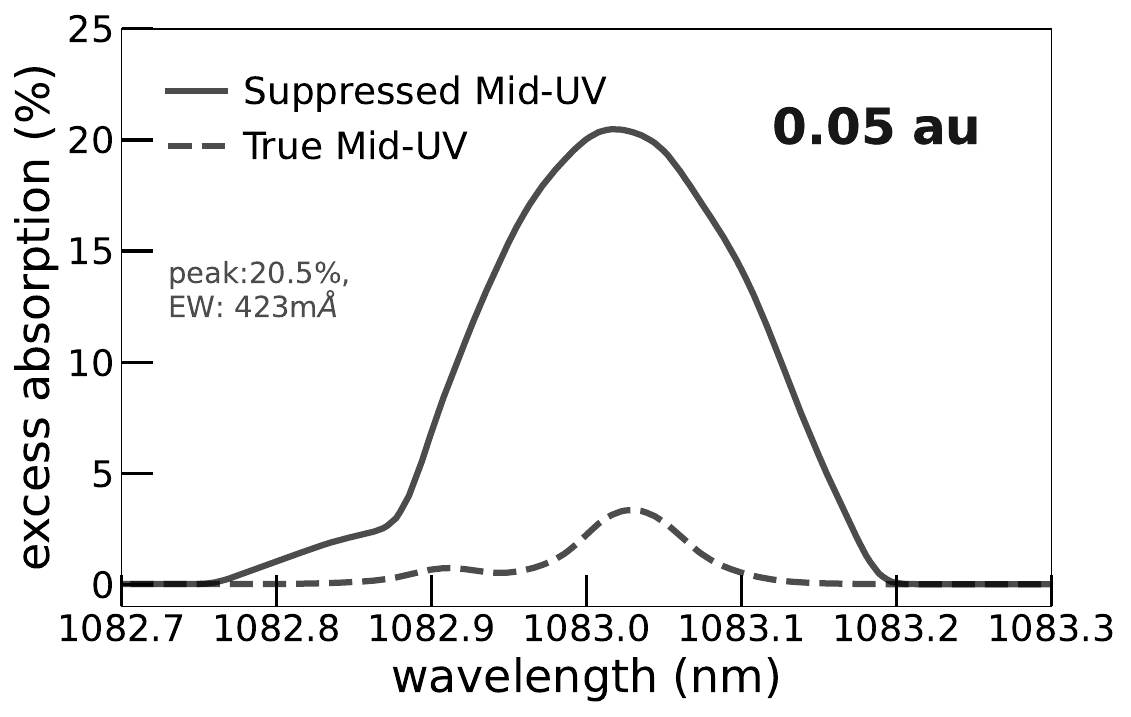} 
  \end{subfigure}
    \caption{ 
        Demonstrative test of how suppressing the mid-UV flux received by the Saturn-mass planet orbiting $\iota$~Hor by a factor of 1000 affects the resulting $\ion{He}{i}$~1083\,nm transit signature. The figure set-up follows that in figure \ref{fig:triplet_mid_UV_supressed_Sun}. 
         }
 \label{fig:triplet_mid_UV_supressed_Iota}
\end{figure}

Figure \ref{fig:triplet_IotaHor} displays the predicted $\ion{He}{i}$~1083\,nm transit signatures obtained using the constructed activity maximum SED of $\iota$~Hor presented previously in section \ref{sec:SED_iHor}. Table \ref{Tab:escape_predictions_IotaHor_all} lists the hydrodynamic escape predictions for the same planets. Due to the similar EUV fluxes between the activity cycle minimum and maximum, our model predicts identical hydrodynamic and $\ion{He}{i}$~1083\,nm solutions, hence we do not compare predictions for activity maximum and minimum as done previously for the solar-like case in sections \ref{sec:solar-like_cycle_hydro_effects} and \ref{sec:effects_solar_cycle_helium_sig}. 
Given the higher level of EUV radiation received by the planet (see table \ref{table:Star_lum_min_max}), the atmospheric escape of our modelled planets orbiting $\iota$~Hor is stronger than those orbiting the Sun. This is seen in their larger mass-loss rates and faster terminal velocities given in Table \ref{Tab:escape_predictions_IotaHor_all}. 
For all modelled planets bar the sub-Neptune orbiting at 0.025\,au, this stronger atmospheric escape yields larger equivalent widths for the $\ion{He}{i}$~1083\,nm signatures. This is due mostly to their significantly broader features arising from their faster atmospheric outflows. Despite their stronger atmospheric escape, the $\ion{He}{i}$~1083\,nm excess absorption for the sub-Neptune planets at both orbital distances and the Neptune-mass planet at 0.025\,au is slightly below that found for the activity maximum solar case. The larger radius of 1.185\(R_\odot\) assumed for $\iota$~Hor is likely responsible, having a reducing effect on the predicted $\ion{He}{i}$~1083\,nm transit signature compared to when assuming a smaller solar-sized stellar disk. This will particularly affect weaker $\ion{He}{i}$~1083\,nm signatures with less atmospheric \ion{He}{I}~($2^3$S) material obscuring the stellar disk. The $\sim$5 times larger mid-UV flux of $\iota$~Hor compared to the solar activity maximum will also contribute to weakening the $\iota$~Hor transit signatures, through raising the rate of photoionisations out of the $2^3$S state.

\begin{table}
\caption{Hydrodynamic escape predictions for planets orbiting $\iota$~Hor. The `maximum' and `minimum' activity phases presented in section \ref{sec:SED_iHor} resulted in the same hydrodynamic predictions and hence are not distinguished here. 
}
\label{Tab:escape_predictions_IotaHor_all}
\begin{tabular}{lllll}
\toprule
 & sub-Nep & Nep-mass & Sat-mass & HJ \\ 
  \hline 
\multicolumn{5}{|c|}{\textbf{0.025\,au}} \\ \hline
$\dot{m}$~g/s & 2.2e+13 & 1e+13 & 5.4e+12 & 3.7e+12 \\
$v_{\infty}$~km/s & 86 & 84 & 82 & 79 \\
$T_{\text{peak}}$~K & 11225 & 11244 & 11262 & 11435 \\
 \hline
\multicolumn{5}{|c|}{\textbf{0.05\,au}} \\ \hline
$\dot{m}$~g/s & 1.2e+13 & 4.7e+12 & 2e+12 & 9.8e+11 \\
$v_{\infty}$~km/s & 45 & 42 & 36 & 32 \\
$T_{\text{peak}}$~K & 11016 & 10899 & 10832 & 10971 \\
\bottomrule \end{tabular}
\end{table}

The non-monotonic behaviour of the predicted  $\ion{He}{i}$~1083\,nm signature with orbital distance discussed previously in section \ref{sec:effects_solar_cycle_helium_sig} is even more pronounced for planets orbiting $\iota$~Hor as seen in Figure \ref{fig:triplet_IotaHor}. The predicted $\ion{He}{i}$~1083\,nm equivalent widths for the further 0.05\,au orbiting planets are comparable to that of their closer orbiting 0.025\,au counterparts despite their weaker atmospheric escapes.
It was shown previously in Figure \ref{fig:triplet_mid_UV_supressed_Sun} that the weaker mid-UV flux at further orbital distance can cause unusual behaviour $\ion{He}{i}$~1083\,nm absorption with orbital distance, due to a reduced number of \ion{He}{I}~($2^3$S) state photoionisations. Figure \ref{fig:triplet_mid_UV_supressed_Iota} presents a similar mid-UV suppression test, now assuming $\iota~$Hor as the stellar host. The lowered mid-UV flux $\ion{He}{i}$~1083\,nm profiles are again substantially stronger. In particular, they are significantly broader in the line wings due to absorption by \ion{He}{I}~($2^3$S) in higher-up, faster-moving atmospheric material which otherwise would have been photoionised by mid-UV. While suppressing the mid-UV flux of $\iota~$Hor causes the $\ion{He}{i}$~1083\,nm equivalent width to fall with increasing orbital distance, the peak of the signature still does not fall with orbital distance. Hence, the mid-UV alone is not responsible for the cases of higher $\ion{He}{i}$~1083\,nm peaks at the further orbital distance of 0.05\,au despite the weaker escape seen in figure \ref{fig:triplet_IotaHor}. The drop in predicted mass-loss rates from an orbital distance of 0.025\,au to 0.05\,au is less for planets orbiting $\iota~$Hor compared to the Sun. This is a result of the stronger XUV emission of $\iota~$Hor causing the atmospheric escape for the closer orbiting models at 0.025\,au to be further from the more efficient energy-limited regime of atmospheric escape than the models orbiting the solar-like star at the same distance. This  weaker relative fall-off in atmospheric escape with orbital distance for $\iota~$Hor orbiting planets complicates the behaviour of their $\ion{He}{i}$~1083\,nm transit signatures with orbital distance, compared to the planets orbiting the Sun-like star with atmospheric escape rates and $\ion{He}{i}$~1083\,nm signatures that are more sensitive to the orbital distance. Furthermore, as emphasised in \citet{Biassoni_2024}, the physics driving the strength of the $\ion{He}{i}$~1083\,nm transit signature is far from straightforward. The signature is reliant on the population of the \ion{He}{I}~($2^3$S) state which itself is largely dependant on the mid-UV flux as discussed, in addition to the atmospheric temperature profile, the availability of \ion{He}{+}\footnote{required to populate \ion{He}{I}~($2^3$S) via recombinations} and free electrons\footnote{contributing to both populating \ion{He}{I}~($2^3$S) via recombinations and depopulating via collisions} and hence the XUV flux indirectly. 

\section{Discussion}
\label{sec:discussion}

\subsection{A stellar XUV cycle can cause conflicting $\ion{He}{i}$~1083\,nm detections and non-detections of atmospheric escape }
For both host stars, the $\ion{He}{i}$~1083\,nm profiles of the sub-Neptune and Neptune-mass planets are quite weak, contrary to their predicted mass-loss rates exceeding that of the Saturn-mass and HJ planets given in tables \ref{Tab:escape_predictions_solar_all} and \ref{Tab:escape_predictions_IotaHor_all}. Their comparatively weaker $\ion{He}{i}$~1083\,nm signatures despite their stronger atmospheric escape is a result of their smaller planetary radii \citep[see][Figures 4b and 5b]{Allan_Vidotto_2025_young_He}.
During a phase of stellar activity cycle minimum, with their $\ion{He}{i}$~1083\,nm signatures at their weakest, a $\ion{He}{i}$~1083\,nm transit observation may be sufficiently weak so as to be interpreted as a non-detection. It is feasible that such a non-detection would conflict with a positive detection made on a date closer to the maximum of the stellar activity cycle. 
The higher chance of a successful $\ion{He}{i}$~1083\,nm during the maximum phase of a stellar activity cycle is another possible explanation for the conflicting $\ion{He}{i}$~1083\,nm detections and non-detections reported for various planets in the literature; \citep[Wasp-52b][]{Kirk_2022_Wasp52b_Wasp177b, Allart_2023_SPIRou_He_11_gas_giants}, \citep[TOI-2076b][]{Zhang_4_mini_Nep, Gaidos_2023_he_search_Two_200_Myr_old_Planets_TOI1807_TOI2076}, \citep[TOI-1683b][]{Zhang_4_mini_Nep, Orell-Miquel_2024_MOPYS}. While the predicted profiles of the larger Saturn-mass and HJ planets modelled here should be detectable even during activity minimum, their transit signatures could vary substantially with the cycle phase. 
Furthermore, Saturn-mass and HJ planets having with less atmospheric helium than that assumed here or orbiting less XUV active hosts could also yield conflicting detections and non-detections as a result of a sufficiently varying stellar XUV cycle.

\subsection{Comparing our model predictions to those in the literature}
As mentioned in the introduction, \citet{Hazra_2020} previously studied the effect of the Sun's activity cycle on the atmospheric escape of a HJ planet, finding similar general behaviours to those described previously in Section \ref{sec:solar-like_cycle_hydro_effects} for our similar HJ model at 0.05\,au. Their predicted mass-loss rate for a HD209458b-like planet varies from $\sim7.5 \times 10^{10}~$g/s to $\sim1.5 \times 10^{11}~$g/s from similar phases of activity minimum and maximum of solar cycle 24 as considered here. These escape rates are comparable to the $6.2 \times 10^{10}~$g/s and $1.1 \times 10^{11}~$g/s predictions for our similar HJ model at 0.05\,au. \citet{Hazra_2020} additionally demonstrated that a solar-like activity cycle does not significantly affect the hydrogen Ly-$\alpha$ planetary transit signature. However, their excess absorption and equivalent widths predictions for the hydrogen H-$\alpha$ line varies by factors of 2.84 and 3.57 due to the activity cycle, more sensitive than the $\ion{He}{i}$~1083\,nm predictions found for all modelled planets in our work. The high sensitivity of H-$\alpha$ transit signature to stellar XUV cycles is due to the population of neutral hydrogen in the first excited state being highly responsive to changes in the atmospheric temperature. 

Using our 1D models, we also investigated the effects of activity cycles on the hydrogen Ly-$\alpha$ planetary transit signature, similar to what was done in \citet{Hazra_2020}. We confirm their findings in our present study, namely that the larger mass-loss rate during a maximum phase of the stellar activity cycle strengthens the predicted Lyman-$\alpha$ transit absorption, however the resulting variation is likely too weak to be detectable. However, it is important to remember that the wings of the Ly-$\alpha$ line are affected by the interaction with the stellar wind \citep[e.g.][]{Bourrier+2016_GJ436, 2018MNRAS.479.3115V, Carolan_2021_SW}, and this interaction cannot be captured in 1D investigations, requiring multi-dimensional modelling. The solar wind, for example,  varies through the solar cycle, changing its velocity structure from a bimodal distribution at minimum to a more complex behaviour at maximum \citep{2008GeoRL..3518103M}. The variations in the velocity structure of the solar wind are due to changes in the solar magnetic field, evolving from a more dipolar structure at minimum, to a more complex structure at maximum \citep{2012ApJ...757...96D, 2018MNRAS.480..477V}. Analogous to the solar wind, it is expected that the evolution of surface magnetic fields would also lead to an evolution of stellar wind properties \citep[e.g.][]{2016MNRAS.459.1907N} and, therefore, we speculate that different wind properties across different epochs of the cycle could lead to more substantial changes in the wings of the Ly-$\alpha$ line than what can be captured in 1D investigations. This would be an interesting investigation that future 3D models could help elucidate.

In \citet[][section 3.5.1]{Taylor_Koskinen_He_model_applied_to_hd209_2025}, the  effects of applying a solar-like activity cycle to a HJ HD209458b-like planet on the predicted atmospheric escape and corresponding  $\ion{He}{i}$~1083\,nm signature are discussed. The atmospheric mass-loss rates predicted by their model varies from $1.9 \times 10^{10}$g/s at activity minimum to $4.2 \times 10^{10}$g/s at activity maximum. Their predicted $\ion{He}{i}$~1083\,nm transit signature for the maximum phase of the activity cycle is similar to the HJ planet at 0.05\,au in Figure \ref{fig:triplet}, while that for activity minimum is notably weaker, peaking with an excess absorption of only $\sim$0.8 per cent (ours is 3.4 per cent). 
A variety of potential reasons could produce the mentioned differences in predicted atmospheric escape and  $\ion{He}{i}$~1083\,nm signatures, from slight variations in the assumed input parameters (in particular the received XUV and mid-UV fluxes to which both models are highly sensitive) as well as inherent differences in the hydrodynamic modelling approaches and treatment of the \ion{He}{I}~($2^3$S) population. While a detailed comparison of the two modelling approaches is beyond the scope of this paper, both studies agree that a solar-like activity cycle in HD209458 would produce significant variability in the  $\ion{He}{i}$~1083\,nm transit signature of its HJ planet due to strong variations in the planet's atmospheric escape.

\section{Conclusions}
\label{sec:conclusions}

In this work, we explored how stellar activity cycles can affect the atmospheric escape and $\ion{He}{i}$~1083\,nm signature of highly irradiated sub-Neptune, Neptune-mass, Saturn-mass and Hot-Jupiter exoplanets. As described in sections \ref{sec:XUV_cycle_Sun} and \ref{sec:SED_iHor}, we considered two types of stellar activity cycles, first a solar-like activity cycle based on solar cycle 24, selected for its wavelength coverage suitable to our atmospheric escape modelling. We also reconstructed SEDs of the active star $\iota$~Hor during both a phase of activity maximum and minimum. Even though iot Hor shows an X-ray cycle, the resulting small variation in EUV flux does not change the modelled atmospheric escape and $\ion{He}{i}$~1083\,nm signatures. 

We showed in section \ref{sec:solar-like_cycle_hydro_effects} that a solar-like activity cycle can strongly influence the atmospheric escape of highly irradiated exoplanets, with mass-loss rate predictions varying by factors of 1.68--1.98 for a phase of activity maximum compared to minimum. Substantial cyclic variations in the EUV flux are responsible for this activity cycle influence, with these high-energy photons heating the planetary atmospheres and driving their escape. This variable atmospheric escape with stellar activity cycle also impacts the resulting $\ion{He}{i}$~1083\,nm transit signatures, a popular tracer used in detecting exoplanetary atmospheric escape. Section \ref{sec:effects_solar_cycle_helium_sig} explores this, with the solar-like activity cycle causing the predicted $\ion{He}{i}$~1083\,nm signatures to vary by factors of 1.55--2.16 and 1.98--2.42 in their peak excess absorption and equivalent width, respectively. A solar-like activity cycle is also shown to have slightly greater influence on the atmospheric escape and helium signature of further orbiting planets (0.05 compared to 0.025\,au). This is the result of a heightened sensitivity to the received EUV flux at lower irradiation levels, where the atmospheric escape is closer to the more efficient energy-limited regime. 

Despite substantially weaker atmospheric escape at further orbital distances, the $\ion{He}{i}$~1083\,nm predictions of the planets orbiting at 0.05\,au are not significantly weaker than those at 0.025\,au. The behaviour of those planets orbiting the more active $\iota$~Hor star is even more extreme, exhibiting comparable or even slightly stronger $\ion{He}{i}$~1083\,nm features for the more distant planets, as shown in section \ref{sec:results_iota_hor}. This unusual behaviour of the $\ion{He}{i}$~1083\,nm with the orbital distance was previously shown in the study of \citet{Biassoni_2024}. Here, we propose that lower mid-UV fluxes at more distant orbits leading to less photoionisations out of the \ion{He}{I}~($2^3$S) state is causing this behaviour. At more distant orbits, the reduction in depopulation out of the observational important $2^3$S state at least partially counterbalances the $\ion{He}{i}$~1083\,nm reducing effect of weaker atmospheric escapes. 

From an observational standpoint, the $\ion{He}{i}$~1083\,nm behaviour with orbital distance leads to the counter-intuitive result that observational studies could benefit from performing $\ion{He}{i}$~1083\,nm transit spectroscopy for planets at further orbits, despite their weaker atmospheric escape rates. Whether the host star undergoes significant variations in its emitted XUV flux is also an important consideration for future observations. Granted there is sufficient XUV variation, timing a transit observation for a phase of activity maximum would raise the possibility of a successful detection. Stellar activity cycles could also help explain some of the conflicting $\ion{He}{i}$~1083\,nm detections and non-detections for the same planet, with detections being more likely during a phase of activity maximum. While time costly,  detecting more and continual studying of known stellar activity cycles of planet hosts, particularly at XUV wavelengths, would greatly aid the current understanding of how stellar activity cycles are impacting planetary atmospheric escape and its associated signatures.

\section*{Acknowledgements}
We thank the anonymous referee for their helpful feedback which has improved this work. APA and AAV acknowledge funding from the European Research Council (ERC) under the European Union's Horizon 2020 research and innovation programme (grant agreement No 817540, ASTROFLOW). AAV acknowledges funding from the Dutch Research Council (NWO), with project number VI.C.232.041 of the Talent Programme Vici. JSF acknowledges support from the
Agencia Estatal de Investigación del Ministerio de Ciencia e Innovación (AEI-
MCINN) under grant PID2022-137241NB-C42. The authors thank Dr. E. Sandford for discussions on the available solar spectral irradiance data. 
The results presented in this document rely on data measured from the Thermosphere Ionosphere Mesosphere Energetics and Dynamics (TIMED) Solar EUV Experiment (SEE) and are available from the TIMED SEE website at \url{https://lasp.colorado.edu/see/data/}. 
The results presented in this document rely on data measured from the Solar Radiation Climate Experiment (SORCE) and are available at \url{https://lasp.colorado.edu/sorce/data/}. 
The TIMED SEE and SORCE data were accessed via the LASP Interactive Solar Irradiance Datacenter (LISIRD) (\url{https://lasp.colorado.edu/lisird/}).

\section*{Data Availability}
The data described in this article will be shared on reasonable request to the corresponding author. The $\iota$~Hor SED data will be made available on the Strasbourg astronomical Data Centre (CDS) website.

\bibliographystyle{mnras}
\bibliography{example}

\begin{thebibliography}{}
\makeatletter
\relax
\def\mn@urlcharsother{\let\do\@makeother \do\$\do\&\do\#\do\^\do\_\do\%\do\~}
\def\mn@doi{\begingroup\mn@urlcharsother \@ifnextchar [ {\mn@doi@}
  {\mn@doi@[]}}
\def\mn@doi@[#1]#2{\def\@tempa{#1}\ifx\@tempa\@empty \href
  {http://dx.doi.org/#2} {doi:#2}\else \href {http://dx.doi.org/#2} {#1}\fi
  \endgroup}
\def\mn@eprint#1#2{\mn@eprint@#1:#2::\@nil}
\def\mn@eprint@arXiv#1{\href {http://arxiv.org/abs/#1} {{\tt arXiv:#1}}}
\def\mn@eprint@dblp#1{\href {http://dblp.uni-trier.de/rec/bibtex/#1.xml}
  {dblp:#1}}
\def\mn@eprint@#1:#2:#3:#4\@nil{\def\@tempa {#1}\def\@tempb {#2}\def\@tempc
  {#3}\ifx \@tempc \@empty \let \@tempc \@tempb \let \@tempb \@tempa \fi \ifx
  \@tempb \@empty \def\@tempb {arXiv}\fi \@ifundefined
  {mn@eprint@\@tempb}{\@tempb:\@tempc}{\expandafter \expandafter \csname
  mn@eprint@\@tempb\endcsname \expandafter{\@tempc}}}

\bibitem[\protect\citeauthoryear{{Alam} et~al.,}{{Alam}
  et~al.}{2024}]{Alam_2024_young_non_detects}
{Alam} M.~K.,  et~al., 2024, \mn@doi [\aj] {10.3847/1538-3881/ad50d4}, \href
  {https://ui.adsabs.harvard.edu/abs/2024AJ....168..102A} {168, 102}

\bibitem[\protect\citeauthoryear{{Allan} \& {Vidotto}}{{Allan} \&
  {Vidotto}}{2019}]{Allan_Vidotto_2019}
{Allan} A.~P.,  {Vidotto} A.~A.,  2019, \mn@doi [\mnras]
  {10.1093/mnras/stz2842}, \href
  {https://ui.adsabs.harvard.edu/abs/2019MNRAS.490.3760A} {490, 3760}

\bibitem[\protect\citeauthoryear{{Allan} \& {Vidotto}}{{Allan} \&
  {Vidotto}}{2025}]{Allan_Vidotto_2025_young_He}
{Allan} A.~P.,  {Vidotto} A.~A.,  2025, \mn@doi [\mnras]
  {10.1093/mnras/staf566}, \href
  {https://ui.adsabs.harvard.edu/abs/2025MNRAS.539.2144A} {539, 2144}

\bibitem[\protect\citeauthoryear{{Allan}, {Vidotto}, {Villarreal D'Angelo},
  {Dos Santos}  \& {Driessen}}{{Allan} et~al.}{2024}]{Allan_et_al_He_evol_2024}
{Allan} A.~P.,  {Vidotto} A.~A.,  {Villarreal D'Angelo} C.,  {Dos Santos}
  L.~A.,   {Driessen} F.~A.,  2024, \mn@doi [\mnras] {10.1093/mnras/stad3432},
  \href {https://ui.adsabs.harvard.edu/abs/2024MNRAS.527.4657A} {527, 4657}

\bibitem[\protect\citeauthoryear{{Allart} et~al.,}{{Allart}
  et~al.}{2023}]{Allart_2023_SPIRou_He_11_gas_giants}
{Allart} R.,  et~al., 2023, \mn@doi [\aap] {10.1051/0004-6361/202245832}, \href
  {https://ui.adsabs.harvard.edu/abs/2023A&A...677A.164A} {677, A164}

\bibitem[\protect\citeauthoryear{{Amazo-G{\'o}mez} et~al.,}{{Amazo-G{\'o}mez}
  et~al.}{2023}]{ama23}
{Amazo-G{\'o}mez} E.~M.,  et~al., 2023, \mn@doi [\mnras]
  {10.1093/mnras/stad2086}, \href
  {https://ui.adsabs.harvard.edu/abs/2023MNRAS.524.5725A} {524, 5725}

\bibitem[\protect\citeauthoryear{{Ayres}}{{Ayres}}{2020}]{Ayres2020_Alpha_Centauri_AandB_Xray_cycle}
{Ayres} T.~R.,  2020, \mn@doi [\apjs] {10.3847/1538-4365/aba3c6}, \href
  {https://ui.adsabs.harvard.edu/abs/2020ApJS..250...16A} {250, 16}

\bibitem[\protect\citeauthoryear{{Ayres}}{{Ayres}}{2025}]{Ayres_2025_Landscape_Coronal_X-Ray_Var_and_Cycles}
{Ayres} T.,  2025, \mn@doi [\aj] {10.3847/1538-3881/adc570}, \href
  {https://ui.adsabs.harvard.edu/abs/2025AJ....169..281A} {169, 281}

\bibitem[\protect\citeauthoryear{{Baliunas} et~al.,}{{Baliunas}
  et~al.}{1995}]{Baliunas_1995_mount_wilson_project2}
{Baliunas} S.~L.,  et~al., 1995, \mn@doi [\apj] {10.1086/175072}, \href
  {https://ui.adsabs.harvard.edu/abs/1995ApJ...438..269B} {438, 269}

\bibitem[\protect\citeauthoryear{{Biassoni}, {Caldiroli}, {Gallo}, {Haardt},
  {Spinelli}  \& {Borsa}}{{Biassoni} et~al.}{2024}]{Biassoni_2024}
{Biassoni} F.,  {Caldiroli} A.,  {Gallo} E.,  {Haardt} F.,  {Spinelli} R.,
  {Borsa} F.,  2024, \mn@doi [\aap] {10.1051/0004-6361/202347517}, \href
  {https://ui.adsabs.harvard.edu/abs/2024A&A...682A.115B} {682, A115}

\bibitem[\protect\citeauthoryear{{Bonomo} et~al.,}{{Bonomo}
  et~al.}{2017}]{Bonomo_2017A&A...602A.107B}
{Bonomo} A.~S.,  et~al., 2017, \mn@doi [\aap] {10.1051/0004-6361/201629882},
  \href {https://ui.adsabs.harvard.edu/abs/2017A&A...602A.107B} {602, A107}

\bibitem[\protect\citeauthoryear{{Boro Saikia} et~al.,}{{Boro Saikia}
  et~al.}{2018}]{Sudeshna_Boro_Saikia_2018_chromospheric_activity_catalogue}
{Boro Saikia} S.,  et~al., 2018, \mn@doi [\aap] {10.1051/0004-6361/201629518},
  \href {https://ui.adsabs.harvard.edu/abs/2018A&A...616A.108B} {616, A108}

\bibitem[\protect\citeauthoryear{{Bourrier}, {Lecavelier des Etangs},
  {Ehrenreich}, {Tanaka}  \& {Vidotto}}{{Bourrier}
  et~al.}{2016}]{Bourrier+2016_GJ436}
{Bourrier} V.,  {Lecavelier des Etangs} A.,  {Ehrenreich} D.,  {Tanaka} Y.~A.,
   {Vidotto} A.~A.,  2016, \mn@doi [\aap] {10.1051/0004-6361/201628362}, \href
  {https://ui.adsabs.harvard.edu/abs/2016A&A...591A.121B} {591, A121}

\bibitem[\protect\citeauthoryear{{Caldiroli}, {Haardt}, {Gallo}, {Spinelli},
  {Malsky}  \& {Rauscher}}{{Caldiroli} et~al.}{2021}]{Caldiroli_2021_ATES}
{Caldiroli} A.,  {Haardt} F.,  {Gallo} E.,  {Spinelli} R.,  {Malsky} I.,
  {Rauscher} E.,  2021, \mn@doi [\aap] {10.1051/0004-6361/202141497}, \href
  {https://ui.adsabs.harvard.edu/abs/2021A&A...655A..30C} {655, A30}

\bibitem[\protect\citeauthoryear{{Caldiroli}, {Haardt}, {Gallo}, {Spinelli},
  {Malsky}  \& {Rauscher}}{{Caldiroli} et~al.}{2022}]{Caldiroli_2022_evap_eff}
{Caldiroli} A.,  {Haardt} F.,  {Gallo} E.,  {Spinelli} R.,  {Malsky} I.,
  {Rauscher} E.,  2022, \mn@doi [\aap] {10.1051/0004-6361/202142763}, \href
  {https://ui.adsabs.harvard.edu/abs/2022A&A...663A.122C} {663, A122}

\bibitem[\protect\citeauthoryear{{Carolan}, {Vidotto}, {Villarreal D'Angelo}
  \& {Hazra}}{{Carolan} et~al.}{2021}]{Carolan_2021_SW}
{Carolan} S.,  {Vidotto} A.~A.,  {Villarreal D'Angelo} C.,   {Hazra} G.,  2021,
  \mn@doi [\mnras] {10.1093/mnras/staa3431}, \href
  {https://ui.adsabs.harvard.edu/abs/2021MNRAS.500.3382C} {500, 3382}

\bibitem[\protect\citeauthoryear{{Castelli} \& {Kurucz}}{{Castelli} \&
  {Kurucz}}{2003}]{Castelli_Kurucz_2003_new_grid_ATLAS9}
{Castelli} F.,  {Kurucz} R.~L.,  2003, in {Piskunov} N.,  {Weiss} W.~W.,
  {Gray} D.~F.,  eds,  IAU Symposium Vol. 210, Modelling of Stellar
  Atmospheres. p.~A20 (\mn@eprint {arXiv} {astro-ph/0405087}),
  \mn@doi{10.48550/arXiv.astro-ph/0405087}

\bibitem[\protect\citeauthoryear{{Cauley}, {Kuckein}, {Redfield}, {Shkolnik},
  {Denker}, {Llama}  \& {Verma}}{{Cauley}
  et~al.}{2018}]{Cauley_2018_effects_stellar_activity_exo_transit}
{Cauley} P.~W.,  {Kuckein} C.,  {Redfield} S.,  {Shkolnik} E.~L.,  {Denker} C.,
   {Llama} J.,   {Verma} M.,  2018, \mn@doi [\aj] {10.3847/1538-3881/aaddf9},
  \href {https://ui.adsabs.harvard.edu/abs/2018AJ....156..189C} {156, 189}

\bibitem[\protect\citeauthoryear{{Clette}, {Svalgaard}, {Vaquero}  \&
  {Cliver}}{{Clette} et~al.}{2014}]{Clette_2014_sunspot_number_400years}
{Clette} F.,  {Svalgaard} L.,  {Vaquero} J.~M.,   {Cliver} E.~W.,  2014,
  \mn@doi [\ssr] {10.1007/s11214-014-0074-2}, \href
  {https://ui.adsabs.harvard.edu/abs/2014SSRv..186...35C} {186, 35}

\bibitem[\protect\citeauthoryear{{Coffaro} et~al.,}{{Coffaro}
  et~al.}{2020}]{Coffaro_2020_eps_eri_xray_cycle}
{Coffaro} M.,  et~al., 2020, \mn@doi [\aap] {10.1051/0004-6361/201936479},
  \href {https://ui.adsabs.harvard.edu/abs/2020A&A...636A..49C} {636, A49}

\bibitem[\protect\citeauthoryear{{DeRosa}, {Brun}  \& {Hoeksema}}{{DeRosa}
  et~al.}{2012}]{2012ApJ...757...96D}
{DeRosa} M.~L.,  {Brun} A.~S.,   {Hoeksema} J.~T.,  2012, \mn@doi [\apj]
  {10.1088/0004-637X/757/1/96}, \href
  {https://ui.adsabs.harvard.edu/abs/2012ApJ...757...96D} {757, 96}

\bibitem[\protect\citeauthoryear{{Dos Santos}}{{Dos
  Santos}}{2023}]{Dos_Santos_2022_iauga_obs_of_pl_winds_outflows}
{Dos Santos} L.~A.,  2023, in {Vidotto} A.~A.,  {Fossati} L.,   {Vink} J.~S.,
  eds,  IAU Symposium Vol. 370, Winds of Stars and Exoplanets. pp 56--71
  (\mn@eprint {arXiv} {2211.16243}), \mn@doi{10.1017/S1743921322004239}

\bibitem[\protect\citeauthoryear{{Drake}}{{Drake}}{1971}]{Drake1971}
{Drake} G.~W.,  1971, \mn@doi [\pra] {10.1103/PhysRevA.3.908}, \href
  {https://ui.adsabs.harvard.edu/abs/1971PhRvA...3..908D} {3, 908}

\bibitem[\protect\citeauthoryear{{Egeland}, {Soon}, {Baliunas}, {Hall},
  {Pevtsov}  \& {Bertello}}{{Egeland}
  et~al.}{2017}]{Egeland_2017_S_index_of_sun}
{Egeland} R.,  {Soon} W.,  {Baliunas} S.,  {Hall} J.~C.,  {Pevtsov} A.~A.,
  {Bertello} L.,  2017, \mn@doi [\apj] {10.3847/1538-4357/835/1/25}, \href
  {https://ui.adsabs.harvard.edu/abs/2017ApJ...835...25E} {835, 25}

\bibitem[\protect\citeauthoryear{{Favata}, {Micela}, {Orlando}, {Schmitt},
  {Sciortino}  \& {Hall}}{{Favata}
  et~al.}{2008}]{Favata_2008_HD81809_xray_cycle}
{Favata} F.,  {Micela} G.,  {Orlando} S.,  {Schmitt} J.~H.~M.~M.,  {Sciortino}
  S.,   {Hall} J.,  2008, \mn@doi [\aap] {10.1051/0004-6361:200809694}, \href
  {https://ui.adsabs.harvard.edu/abs/2008A&A...490.1121F} {490, 1121}

\bibitem[\protect\citeauthoryear{{Fortney} \& {Nettelmann}}{{Fortney} \&
  {Nettelmann}}{2010}]{Fortney_and_Nettelmann2010}
{Fortney} J.~J.,  {Nettelmann} N.,  2010, \mn@doi [\ssr]
  {10.1007/s11214-009-9582-x}, \href
  {https://ui.adsabs.harvard.edu/abs/2010SSRv..152..423F} {152, 423}

\bibitem[\protect\citeauthoryear{{France} et~al.,}{{France}
  et~al.}{2016}]{France_2016}
{France} K.,  et~al., 2016, \mn@doi [\apj] {10.3847/0004-637X/820/2/89}, \href
  {https://ui.adsabs.harvard.edu/abs/2016ApJ...820...89F} {820, 89}

\bibitem[\protect\citeauthoryear{{Fuhrmeister} et~al.,}{{Fuhrmeister}
  et~al.}{2020}]{Fuhrmeister_2020_Carmenes_Mdwarfs_variability_triplet}
{Fuhrmeister} B.,  et~al., 2020, \mn@doi [\aap] {10.1051/0004-6361/202038279},
  \href {https://ui.adsabs.harvard.edu/abs/2020A&A...640A..52F} {640, A52}

\bibitem[\protect\citeauthoryear{{Gaidos} et~al.,}{{Gaidos}
  et~al.}{2023}]{Gaidos_2023_he_search_Two_200_Myr_old_Planets_TOI1807_TOI2076}
{Gaidos} E.,  et~al., 2023, \mn@doi [\mnras] {10.1093/mnras/stac3301}, \href
  {https://ui.adsabs.harvard.edu/abs/2023MNRAS.518.3777G} {518, 3777}

\bibitem[\protect\citeauthoryear{{Guilluy} et~al.,}{{Guilluy}
  et~al.}{2020}]{2020_Guilluy_GAPS_He_189}
{Guilluy} G.,  et~al., 2020, \mn@doi [\aap] {10.1051/0004-6361/202037644},
  \href {https://ui.adsabs.harvard.edu/abs/2020A&A...639A..49G} {639, A49}

\bibitem[\protect\citeauthoryear{{Hall}, {Lockwood}  \& {Skiff}}{{Hall}
  et~al.}{2007}]{Hall_2007_Lowell_observatory}
{Hall} J.~C.,  {Lockwood} G.~W.,   {Skiff} B.~A.,  2007, \mn@doi [\aj]
  {10.1086/510356}, \href
  {https://ui.adsabs.harvard.edu/abs/2007AJ....133..862H} {133, 862}

\bibitem[\protect\citeauthoryear{{Hathaway}}{{Hathaway}}{2010}]{2010LRSP....7....1H}
{Hathaway} D.~H.,  2010, \mn@doi [Living Reviews in Solar Physics]
  {10.12942/lrsp-2010-1}, \href
  {https://ui.adsabs.harvard.edu/abs/2010LRSP....7....1H} {7, 1}

\bibitem[\protect\citeauthoryear{{Hazra}, {Vidotto}  \& {D'Angelo}}{{Hazra}
  et~al.}{2020}]{Hazra_2020}
{Hazra} G.,  {Vidotto} A.~A.,   {D'Angelo} C.~V.,  2020, \mn@doi [\mnras]
  {10.1093/mnras/staa1815}, \href
  {https://ui.adsabs.harvard.edu/abs/2020MNRAS.496.4017H} {496, 4017}

\bibitem[\protect\citeauthoryear{{Hempelmann}, {Robrade}, {Schmitt}, {Favata},
  {Baliunas}  \& {Hall}}{{Hempelmann}
  et~al.}{2006}]{Hempelmann_2006_61Cygni_xraycycle}
{Hempelmann} A.,  {Robrade} J.,  {Schmitt} J.~H.~M.~M.,  {Favata} F.,
  {Baliunas} S.~L.,   {Hall} J.~C.,  2006, \mn@doi [\aap]
  {10.1051/0004-6361:20065459}, \href
  {https://ui.adsabs.harvard.edu/abs/2006A&A...460..261H} {460, 261}

\bibitem[\protect\citeauthoryear{{Johnstone}, {Bartel}  \&
  {G{\"u}del}}{{Johnstone} et~al.}{2021}]{Johnstone_2021}
{Johnstone} C.~P.,  {Bartel} M.,   {G{\"u}del} M.,  2021, \mn@doi [\aap]
  {10.1051/0004-6361/202038407}, \href
  {https://ui.adsabs.harvard.edu/abs/2021A&A...649A..96J} {649, A96}

\bibitem[\protect\citeauthoryear{{Kirk}, {Dos Santos}, {L{\'o}pez-Morales},
  {Alam}, {Oklop{\v{c}}i{\'c}}, {MacLeod}, {Zeng}  \& {Zhou}}{{Kirk}
  et~al.}{2022}]{Kirk_2022_Wasp52b_Wasp177b}
{Kirk} J.,  {Dos Santos} L.~A.,  {L{\'o}pez-Morales} M.,  {Alam} M.~K.,
  {Oklop{\v{c}}i{\'c}} A.,  {MacLeod} M.,  {Zeng} L.,   {Zhou} G.,  2022,
  \mn@doi [\aj] {10.3847/1538-3881/ac722f}, \href
  {https://ui.adsabs.harvard.edu/abs/2022AJ....164...24K} {164, 24}

\bibitem[\protect\citeauthoryear{{Krishnamurthy} \& {Cowan}}{{Krishnamurthy} \&
  {Cowan}}{2024}]{Krishnamurthy_Cowan_2024}
{Krishnamurthy} V.,  {Cowan} N.~B.,  2024, \mn@doi [\aj]
  {10.3847/1538-3881/ad5441}, \href
  {https://ui.adsabs.harvard.edu/abs/2024AJ....168...30K} {168, 30}

\bibitem[\protect\citeauthoryear{{Krolikowski}, {Kraus}, {Tofflemire},
  {Morley}, {Mann}  \& {Vanderburg}}{{Krolikowski}
  et~al.}{2024}]{krolikowski2024strength}
{Krolikowski} D.~M.,  {Kraus} A.~L.,  {Tofflemire} B.~M.,  {Morley} C.~V.,
  {Mann} A.~W.,   {Vanderburg} A.,  2024, \mn@doi [\aj]
  {10.3847/1538-3881/ad0f22}, \href
  {https://ui.adsabs.harvard.edu/abs/2024AJ....167...79K} {167, 79}

\bibitem[\protect\citeauthoryear{{Kuckein}, {Collados}  \& {Manso
  Sainz}}{{Kuckein} et~al.}{2015}]{Kuckein_2015_solar_flare}
{Kuckein} C.,  {Collados} M.,   {Manso Sainz} R.,  2015, \mn@doi [\apjl]
  {10.1088/2041-8205/799/2/L25}, \href
  {https://ui.adsabs.harvard.edu/abs/2015ApJ...799L..25K} {799, L25}

\bibitem[\protect\citeauthoryear{{Mahadevan} et~al.,}{{Mahadevan}
  et~al.}{2012}]{Mahadevan_2012_HPF}
{Mahadevan} S.,  et~al., 2012, in {McLean} I.~S.,  {Ramsay} S.~K.,   {Takami}
  H.,  eds,  Society of Photo-Optical Instrumentation Engineers (SPIE)
  Conference Series Vol. 8446, Ground-based and Airborne Instrumentation for
  Astronomy IV. p. 84461S (\mn@eprint {arXiv} {1209.1686}),
  \mn@doi{10.1117/12.926102}

\bibitem[\protect\citeauthoryear{{McComas}, {Ebert}, {Elliott}, {Goldstein},
  {Gosling}, {Schwadron}  \& {Skoug}}{{McComas}
  et~al.}{2008}]{2008GeoRL..3518103M}
{McComas} D.~J.,  {Ebert} R.~W.,  {Elliott} H.~A.,  {Goldstein} B.~E.,
  {Gosling} J.~T.,  {Schwadron} N.~A.,   {Skoug} R.~M.,  2008, \mn@doi [\grl]
  {10.1029/2008GL034896}, \href
  {https://ui.adsabs.harvard.edu/abs/2008GeoRL..3518103M} {35, L18103}

\bibitem[\protect\citeauthoryear{{McCreery}, {Dos Santos}, {Espinoza}, {Allart}
   \& {Kirk}}{{McCreery} et~al.}{2025}]{McCreery_2025}
{McCreery} P.,  {Dos Santos} L.~A.,  {Espinoza} N.,  {Allart} R.,   {Kirk} J.,
  2025, \mn@doi [\apj] {10.3847/1538-4357/ada6b9}, \href
  {https://ui.adsabs.harvard.edu/abs/2025ApJ...980..125M} {980, 125}

\bibitem[\protect\citeauthoryear{{Mercier} et~al.,}{{Mercier}
  et~al.}{2025}]{Mercier_2025_var_He_triplet_detection_lims_evap_atm}
{Mercier} S.~J.,  et~al., 2025, arXiv e-prints, \href
  {https://ui.adsabs.harvard.edu/abs/2025arXiv250721290M} {p. arXiv:2507.21290}

\bibitem[\protect\citeauthoryear{{Metcalfe}, {Basu}, {Henry}, {Soderblom},
  {Judge}, {Kn{\"o}lker}, {Mathur}  \& {Rempel}}{{Metcalfe}
  et~al.}{2010}]{Metcalfe_2010}
{Metcalfe} T.~S.,  {Basu} S.,  {Henry} T.~J.,  {Soderblom} D.~R.,  {Judge}
  P.~G.,  {Kn{\"o}lker} M.,  {Mathur} S.,   {Rempel} M.,  2010, \mn@doi [\apjl]
  {10.1088/2041-8205/723/2/L213}, \href
  {https://ui.adsabs.harvard.edu/abs/2010ApJ...723L.213M} {723, L213}

\bibitem[\protect\citeauthoryear{{Mittag}, {Robrade}, {Schmitt}, {Hempelmann},
  {Gonz{\'a}lez-P{\'e}rez}  \& {Schr{\"o}der}}{{Mittag}
  et~al.}{2017}]{tau_boo_xray_cycle2017}
{Mittag} M.,  {Robrade} J.,  {Schmitt} J.~H.~M.~M.,  {Hempelmann} A.,
  {Gonz{\'a}lez-P{\'e}rez} J.~N.,   {Schr{\"o}der} K.~P.,  2017, \mn@doi [\aap]
  {10.1051/0004-6361/201629156}, \href
  {https://ui.adsabs.harvard.edu/abs/2017A&A...600A.119M} {600, A119}

\bibitem[\protect\citeauthoryear{Murray-Clay, Chiang  \& Murray}{Murray-Clay
  et~al.}{2009}]{Murray-Clay2009}
Murray-Clay R.~A.,  Chiang E.~I.,   Murray N.,  2009, \mn@doi [Astrophysical
  Journal] {10.1088/0004-637X/693/1/23}, 693, 23

\bibitem[\protect\citeauthoryear{{Nicholson} et~al.,}{{Nicholson}
  et~al.}{2016}]{2016MNRAS.459.1907N}
{Nicholson} B.~A.,  et~al., 2016, \mn@doi [\mnras] {10.1093/mnras/stw731},
  \href {https://ui.adsabs.harvard.edu/abs/2016MNRAS.459.1907N} {459, 1907}

\bibitem[\protect\citeauthoryear{{Oklop{\v{c}}i{\'c}}}{{Oklop{\v{c}}i{\'c}}}{2019}]{Oklopcic_2019_dep_st_rad}
{Oklop{\v{c}}i{\'c}} A.,  2019, \mn@doi [\apj] {10.3847/1538-4357/ab2f7f},
  \href {https://ui.adsabs.harvard.edu/abs/2019ApJ...881..133O} {881, 133}

\bibitem[\protect\citeauthoryear{{Ol{\'a}h}, {K{\H{o}}v{\'a}ri}, {Petrovay},
  {Soon}, {Baliunas}, {Koll{\'a}th}  \& {Vida}}{{Ol{\'a}h}
  et~al.}{2016}]{Olah_2016_Magnetic_cycles_at_different_ages_of_stars}
{Ol{\'a}h} K.,  {K{\H{o}}v{\'a}ri} Z.,  {Petrovay} K.,  {Soon} W.,  {Baliunas}
  S.,  {Koll{\'a}th} Z.,   {Vida} K.,  2016, \mn@doi [\aap]
  {10.1051/0004-6361/201628479}, \href
  {https://ui.adsabs.harvard.edu/abs/2016A&A...590A.133O} {590, A133}

\bibitem[\protect\citeauthoryear{{Orell-Miquel} et~al.,}{{Orell-Miquel}
  et~al.}{2024}]{Orell-Miquel_2024_MOPYS}
{Orell-Miquel} J.,  et~al., 2024, \mn@doi [\aap] {10.1051/0004-6361/202449411},
  \href {https://ui.adsabs.harvard.edu/abs/2024A&A...689A.179O} {689, A179}

\bibitem[\protect\citeauthoryear{{Orlando}, {Favata}, {Micela}, {Sciortino},
  {Maggio}, {Schmitt}, {Robrade}  \& {Mittag}}{{Orlando}
  et~al.}{2017}]{Orlando_2017_HD81809_xray_cycle}
{Orlando} S.,  {Favata} F.,  {Micela} G.,  {Sciortino} S.,  {Maggio} A.,
  {Schmitt} J.~H.~M.~M.,  {Robrade} J.,   {Mittag} M.,  2017, \mn@doi [\aap]
  {10.1051/0004-6361/201731301}, \href
  {https://ui.adsabs.harvard.edu/abs/2017A&A...605A..19O} {605, A19}

\bibitem[\protect\citeauthoryear{{Owen}}{{Owen}}{2019}]{Owen_EVOL_ATM_ESCAP_REVIEW_2019}
{Owen} J.~E.,  2019, \mn@doi [Annual Review of Earth and Planetary Sciences]
  {10.1146/annurev-earth-053018-060246}, \href
  {https://ui.adsabs.harvard.edu/abs/2019AREPS..47...67O} {47, 67}

\bibitem[\protect\citeauthoryear{{Owen} \& {Alvarez}}{{Owen} \&
  {Alvarez}}{2016}]{Owen_Alvarez_2016}
{Owen} J.~E.,  {Alvarez} M.~A.,  2016, \mn@doi [\apj]
  {10.3847/0004-637X/816/1/34}, \href
  {https://ui.adsabs.harvard.edu/abs/2016ApJ...816...34O} {816, 34}

\bibitem[\protect\citeauthoryear{{Parmentier} \& {Guillot}}{{Parmentier} \&
  {Guillot}}{2014}]{Parmentier_Guillot_I}
{Parmentier} V.,  {Guillot} T.,  2014, \mn@doi [\aap]
  {10.1051/0004-6361/201322342}, \href
  {https://ui.adsabs.harvard.edu/abs/2014A&A...562A.133P} {562, A133}

\bibitem[\protect\citeauthoryear{{P{\'e}rez-Gonz{\'a}lez}, {Greklek-McKeon},
  {Vissapragada}, {Saidel}, {Knutson}, {Linssen}  \&
  {Oklop{\v{c}}i{\'c}}}{{P{\'e}rez-Gonz{\'a}lez}
  et~al.}{2024}]{Perez_Gonzales_2023_TOI-1268b}
{P{\'e}rez-Gonz{\'a}lez} J.,  {Greklek-McKeon} M.,  {Vissapragada} S.,
  {Saidel} M.,  {Knutson} H.~A.,  {Linssen} D.,   {Oklop{\v{c}}i{\'c}} A.,
  2024, \mn@doi [\aj] {10.3847/1538-3881/ad34b6}, \href
  {https://ui.adsabs.harvard.edu/abs/2024AJ....167..214P} {167, 214}

\bibitem[\protect\citeauthoryear{{Robrade}, {Schmitt}  \& {Favata}}{{Robrade}
  et~al.}{2012}]{Robrade_2012_xray_61CygA_and_alpha_Cen_AB}
{Robrade} J.,  {Schmitt} J.~H.~M.~M.,   {Favata} F.,  2012, \mn@doi [\aap]
  {10.1051/0004-6361/201219046}, \href
  {https://ui.adsabs.harvard.edu/abs/2012A&A...543A..84R} {543, A84}

\bibitem[\protect\citeauthoryear{{Salz} et~al.,}{{Salz}
  et~al.}{2018}]{Salz_2018_hE_HD189733}
{Salz} M.,  et~al., 2018, \mn@doi [\aap] {10.1051/0004-6361/201833694}, \href
  {https://ui.adsabs.harvard.edu/abs/2018A&A...620A..97S} {620, A97}

\bibitem[\protect\citeauthoryear{{Sanz-Forcada} \& {Dupree}}{{Sanz-Forcada} \&
  {Dupree}}{2008}]{Sanz-Forcada_Dupree_2008_active_cool_stars_he_1083}
{Sanz-Forcada} J.,  {Dupree} A.~K.,  2008, \mn@doi [\aap]
  {10.1051/0004-6361:20078501}, \href
  {https://ui.adsabs.harvard.edu/abs/2008A&A...488..715S} {488, 715}

\bibitem[\protect\citeauthoryear{Sanz-Forcada, Micela, Ribas, Pollock, Eiroa,
  Velasco, Solano  \& Garcia-Alvarez}{Sanz-Forcada
  et~al.}{2011}]{Sanz-Forcada2011}
Sanz-Forcada J.,  Micela G.,  Ribas I.,  Pollock a. M.~T.,  Eiroa C.,  Velasco
  A.,  Solano E.,   Garcia-Alvarez D.,  2011, \mn@doi [Astronomy and
  Astrophysics] {10.1051/0004-6361/201116594}, 6, 10

\bibitem[\protect\citeauthoryear{{Sanz-Forcada}, {Stelzer}  \&
  {Metcalfe}}{{Sanz-Forcada}
  et~al.}{2013}]{Sanz_Forcada_2013_EARLIER_iota_hor_xray_cycle}
{Sanz-Forcada} J.,  {Stelzer} B.,   {Metcalfe} T.~S.,  2013, \mn@doi [\aap]
  {10.1051/0004-6361/201321388}, \href
  {https://ui.adsabs.harvard.edu/abs/2013A&A...553L...6S} {553, L6}

\bibitem[\protect\citeauthoryear{{Sanz-Forcada}, {Stelzer}, {Coffaro}, {Raetz}
  \& {Alvarado-G{\'o}mez}}{{Sanz-Forcada}
  et~al.}{2019}]{Sanz_Forcada_2019_iota_hor_xray_cycle}
{Sanz-Forcada} J.,  {Stelzer} B.,  {Coffaro} M.,  {Raetz} S.,
  {Alvarado-G{\'o}mez} J.~D.,  2019, \mn@doi [\aap]
  {10.1051/0004-6361/201935703}, \href
  {https://ui.adsabs.harvard.edu/abs/2019A&A...631A..45S} {631, A45}

\bibitem[\protect\citeauthoryear{{Sanz-Forcada} et~al.,}{{Sanz-Forcada}
  et~al.}{2025}]{Sanz-Forcada_2025_HeI_10830}
{Sanz-Forcada} J.,  et~al., 2025, \mn@doi [\aap] {10.1051/0004-6361/202451680},
  \href {https://ui.adsabs.harvard.edu/abs/2025A&A...693A.285S} {693, A285}

\bibitem[\protect\citeauthoryear{{Singh} \& {Pandey}}{{Singh} \&
  {Pandey}}{2024}]{Singh_2024_AB_Dor_xray-cycle}
{Singh} G.,  {Pandey} J.~C.,  2024, \mn@doi [\apj] {10.3847/1538-4357/ad2f2e},
  \href {https://ui.adsabs.harvard.edu/abs/2024ApJ...966...86S} {966, 86}

\bibitem[\protect\citeauthoryear{{Str{\"u}der} et~al.,}{{Str{\"u}der}
  et~al.}{2001}]{Struder_2001_XMM}
{Str{\"u}der} L.,  et~al., 2001, \mn@doi [\aap] {10.1051/0004-6361:20000066},
  \href {https://ui.adsabs.harvard.edu/abs/2001A&A...365L..18S} {365, L18}

\bibitem[\protect\citeauthoryear{{Taylor}, {Koskinen}, {Argenti}, {Lewis},
  {Huang}, {Arfaux}  \& {Lavvas}}{{Taylor}
  et~al.}{2025}]{Taylor_Koskinen_He_model_applied_to_hd209_2025}
{Taylor} A.~R.,  {Koskinen} T.~T.,  {Argenti} L.,  {Lewis} N.,  {Huang} C.,
  {Arfaux} A.,   {Lavvas} P.,  2025, \mn@doi [arXiv e-prints]
  {10.48550/arXiv.2506.08232}, \href
  {https://ui.adsabs.harvard.edu/abs/2025arXiv250608232T} {p. arXiv:2506.08232}

\bibitem[\protect\citeauthoryear{{Turner} et~al.,}{{Turner}
  et~al.}{2001}]{Turner_2001_XMM}
{Turner} M.~J.~L.,  et~al., 2001, \mn@doi [\aap] {10.1051/0004-6361:20000087},
  \href {https://ui.adsabs.harvard.edu/abs/2001A&A...365L..27T} {365, L27}

\bibitem[\protect\citeauthoryear{{Vauclair}, {Laymand}, {Bouchy}, {Vauclair},
  {Hui Bon Hoa}, {Charpinet}  \& {Bazot}}{{Vauclair}
  et~al.}{2008}]{Vauclair_2008_iota_hor}
{Vauclair} S.,  {Laymand} M.,  {Bouchy} F.,  {Vauclair} G.,  {Hui Bon Hoa} A.,
  {Charpinet} S.,   {Bazot} M.,  2008, \mn@doi [\aap]
  {10.1051/0004-6361:20079342}, \href
  {https://ui.adsabs.harvard.edu/abs/2008A&A...482L...5V} {482, L5}

\bibitem[\protect\citeauthoryear{{Vidotto} et~al.,}{{Vidotto}
  et~al.}{2014}]{Vidotto_2014}
{Vidotto} A.~A.,  et~al., 2014, \mn@doi [\mnras] {10.1093/mnras/stu728}, \href
  {https://ui.adsabs.harvard.edu/abs/2014MNRAS.441.2361V} {441, 2361}

\bibitem[\protect\citeauthoryear{{Vidotto}, {Lehmann}, {Jardine}  \&
  {Pevtsov}}{{Vidotto} et~al.}{2018}]{2018MNRAS.480..477V}
{Vidotto} A.~A.,  {Lehmann} L.~T.,  {Jardine} M.,   {Pevtsov} A.~A.,  2018,
  \mn@doi [\mnras] {10.1093/mnras/sty1926}, \href
  {https://ui.adsabs.harvard.edu/abs/2018MNRAS.480..477V} {480, 477}

\bibitem[\protect\citeauthoryear{{Villarreal D'Angelo}, {Esquivel}, {Schneiter}
   \& {Sgr{\'o}}}{{Villarreal D'Angelo} et~al.}{2018}]{2018MNRAS.479.3115V}
{Villarreal D'Angelo} C.,  {Esquivel} A.,  {Schneiter} M.,   {Sgr{\'o}} M.~A.,
  2018, \mn@doi [\mnras] {10.1093/mnras/sty1544}, \href
  {https://ui.adsabs.harvard.edu/abs/2018MNRAS.479.3115V} {479, 3115}

\bibitem[\protect\citeauthoryear{{Vissapragada} et~al.,}{{Vissapragada}
  et~al.}{2021}]{2021_Vissapragada_search_He_in_V1298_tau_system}
{Vissapragada} S.,  et~al., 2021, \mn@doi [\aj] {10.3847/1538-3881/ac1bb0},
  \href {https://ui.adsabs.harvard.edu/abs/2021AJ....162..222V} {162, 222}

\bibitem[\protect\citeauthoryear{{Wang} \& {Dai}}{{Wang} \&
  {Dai}}{2021}]{Wang_Dai2021_WASP-69b}
{Wang} L.,  {Dai} F.,  2021, \mn@doi [\apj] {10.3847/1538-4357/abf1ee}, \href
  {https://ui.adsabs.harvard.edu/abs/2021ApJ...914...98W} {914, 98}

\bibitem[\protect\citeauthoryear{{Wargelin}, {Saar}, {Irving}, {Slavin},
  {Ratzlaff}  \& {do Nascimento}}{{Wargelin} et~al.}{2024}]{Wargelin_2024}
{Wargelin} B.~J.,  {Saar} S.~H.,  {Irving} Z.~A.,  {Slavin} J.~D.,  {Ratzlaff}
  P.,   {do Nascimento} J.-D.,  2024, \mn@doi [\apj]
  {10.3847/1538-4357/ad8faa}, \href
  {https://ui.adsabs.harvard.edu/abs/2024ApJ...977..144W} {977, 144}

\bibitem[\protect\citeauthoryear{{Wilson}}{{Wilson}}{1978}]{Wilson_1978_mount_wilson_project}
{Wilson} O.~C.,  1978, \mn@doi [\apj] {10.1086/156618}, \href
  {https://ui.adsabs.harvard.edu/abs/1978ApJ...226..379W} {226, 379}

\bibitem[\protect\citeauthoryear{{Woods} et~al.,}{{Woods}
  et~al.}{2005}]{Woods_2005_Solar_EUV_experiment}
{Woods} T.~N.,  et~al., 2005, \mn@doi [Journal of Geophysical Research (Space
  Physics)] {10.1029/2004JA010765}, \href
  {https://ui.adsabs.harvard.edu/abs/2005JGRA..110.1312W} {110, A01312}

\bibitem[\protect\citeauthoryear{{Woods}, {Harder}, {Kopp}, {McCabe},
  {Rottman}, {Ryan}  \& {Snow}}{{Woods}
  et~al.}{2021}]{Woods_SORCE_overview_2021}
{Woods} T.~N.,  {Harder} J.~W.,  {Kopp} G.,  {McCabe} D.,  {Rottman} G.,
  {Ryan} S.,   {Snow} M.,  2021, \mn@doi [\solphys]
  {10.1007/s11207-021-01869-3}, \href
  {https://ui.adsabs.harvard.edu/abs/2021SoPh..296..127W} {296, 127}

\bibitem[\protect\citeauthoryear{{Youngblood} et~al.,}{{Youngblood}
  et~al.}{2016}]{Yongblood_2016_MUSCLES}
{Youngblood} A.,  et~al., 2016, \mn@doi [\apj] {10.3847/0004-637X/824/2/101},
  \href {https://ui.adsabs.harvard.edu/abs/2016ApJ...824..101Y} {824, 101}

\bibitem[\protect\citeauthoryear{{Zhang}, {Cauley}, {Knutson}, {France},
  {Kreidberg}, {Oklop{\v{c}}i{\'c}}, {Redfield}  \& {Shkolnik}}{{Zhang}
  et~al.}{2022}]{Zhang_Cauley_et_al_2022_variable_he_abs_hd189733b}
{Zhang} M.,  {Cauley} P.~W.,  {Knutson} H.~A.,  {France} K.,  {Kreidberg} L.,
  {Oklop{\v{c}}i{\'c}} A.,  {Redfield} S.,   {Shkolnik} E.~L.,  2022, \mn@doi
  [\aj] {10.3847/1538-3881/ac9675}, \href
  {https://ui.adsabs.harvard.edu/abs/2022AJ....164..237Z} {164, 237}

\bibitem[\protect\citeauthoryear{{Zhang} et~al.,}{{Zhang}
  et~al.}{2023a}]{Zhoujian_Zhang_2023_HATP32b}
{Zhang} Z.,  et~al., 2023a, \mn@doi [Science Advances]
  {10.1126/sciadv.adf8736}, \href
  {https://ui.adsabs.harvard.edu/abs/2023SciA....9F8736Z} {9, eadf8736}

\bibitem[\protect\citeauthoryear{{Zhang}, {Knutson}, {Dai}, {Wang}, {Ricker},
  {Schwarz}, {Mann}  \& {Collins}}{{Zhang} et~al.}{2023b}]{Zhang_4_mini_Nep}
{Zhang} M.,  {Knutson} H.~A.,  {Dai} F.,  {Wang} L.,  {Ricker} G.~R.,
  {Schwarz} R.~P.,  {Mann} C.,   {Collins} K.,  2023b, \mn@doi [\aj]
  {10.3847/1538-3881/aca75b}, \href
  {https://ui.adsabs.harvard.edu/abs/2023AJ....165...62Z} {165, 62}

\bibitem[\protect\citeauthoryear{{Zirin}}{{Zirin}}{1975}]{Zirin_1975}
{Zirin} H.,  1975, \mn@doi [\apjl] {10.1086/181849}, \href
  {https://ui.adsabs.harvard.edu/abs/1975ApJ...199L..63Z} {199, L63}

\bibitem[\protect\citeauthoryear{{{\v{S}}ubjak} et~al.,}{{{\v{S}}ubjak}
  et~al.}{2022}]{Subjak_2022_TOI1268b}
{{\v{S}}ubjak} J.,  et~al., 2022, \mn@doi [\aap] {10.1051/0004-6361/202142883},
  \href {https://ui.adsabs.harvard.edu/abs/2022A&A...662A.107S} {662, A107}

\bibitem[\protect\citeauthoryear{{von Braun} et~al.,}{{von Braun}
  et~al.}{2012}]{von_Braun_2012_pl_params}
{von Braun} K.,  et~al., 2012, \mn@doi [\apj] {10.1088/0004-637X/753/2/171},
  \href {https://ui.adsabs.harvard.edu/abs/2012ApJ...753..171V} {753, 171}

\makeatother
\end{thebibliography}

\appendix
\section{The XUV variation over the full solar cycle 24}
\label{appendix:XUV_cycle}

\begin{figure}
    \includegraphics[width=0.475\textwidth]{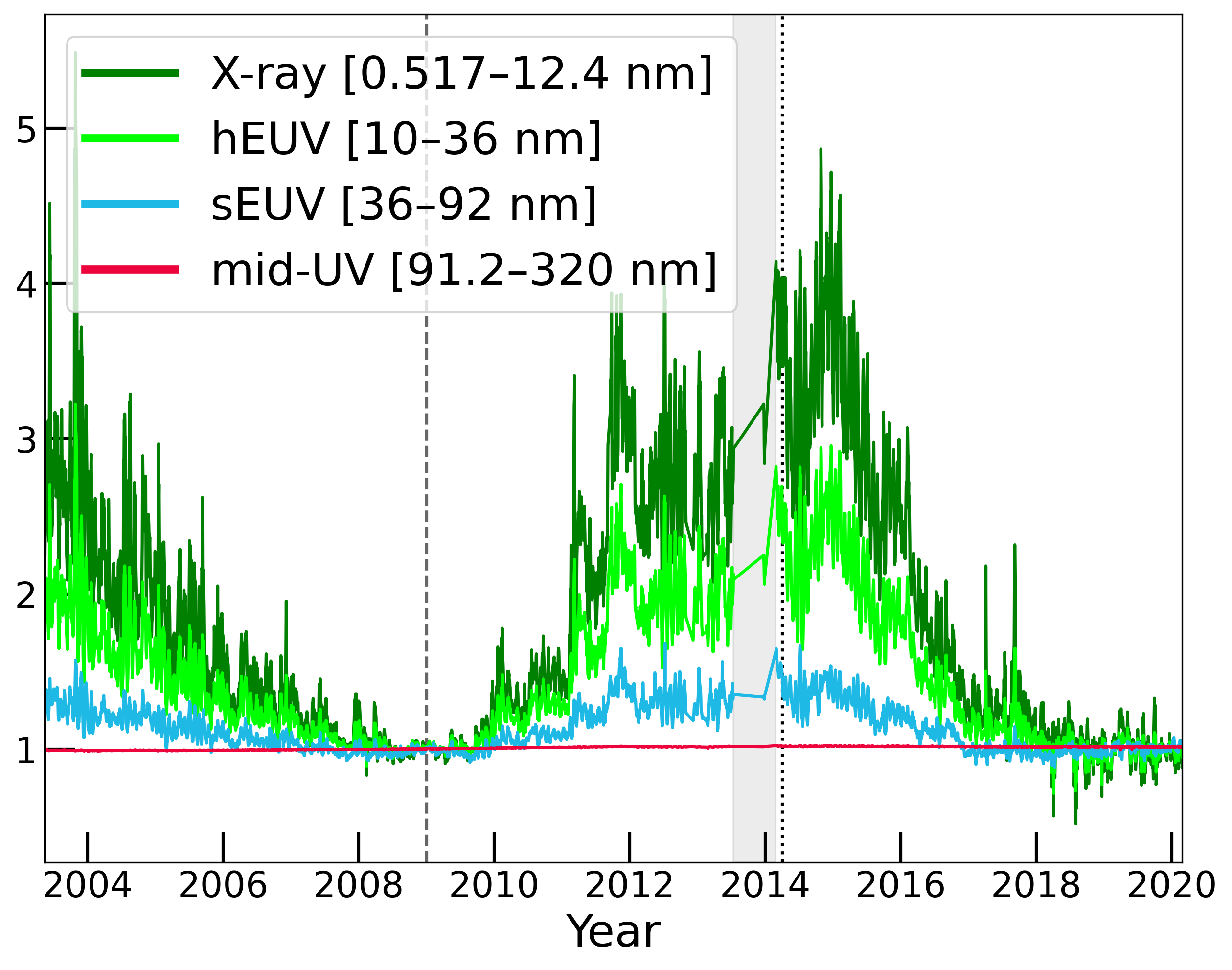} 
    \caption{
      The cyclic nature of the flux of the Sun in our considered wavelength bins. Note that the mid-UV flux shows comparatively negligible cyclic variation. The data used are publicly available and obtained by the SEE instrument of the NASA (TIMED) mission as well as the NASA (SORCE) mission as described in the text. The different colours represent different wavelength bins over which the flux was integrated as indicated by the legend. The plotted values are normalised to the integrated flux within the same wavelength band on the date we assign to an activity minimum phase (vertical dashed line). The vertical dotted line indicates our assumed activity maximum and the grey shaded region indicates a period when there was an issue with the SORCE instruments and hence no available data.
      }
 \label{fig:cycle_integ_fluxes}
\end{figure}

In section \ref{sec:XUV_cycle_Sun}, we describe how we utilise observational solar data in order to obtain fluxes in four different wavelength bins necessary for our model of atmospheric escape. For the sake of completeness, we now show in figure \ref{fig:cycle_integ_fluxes}, the cyclic nature of the emitted high-energy flux over solar cycle 24, with the dates corresponding to the minimum and maximum phases we consider in our modelling marked by the dashed and dotted vertical lines, respectively. 

\bsp	
\label{lastpage}
\end{document}